\documentclass[usenatbib]{mnras}

\usepackage[T1]{fontenc}
\usepackage{ae,aecompl}
\usepackage[dvipsnames]{xcolor}

\usepackage{graphicx}	
\usepackage{amsmath}	
\usepackage{amssymb}	
\usepackage{subfigure}
\usepackage{soul}
\usepackage{cancel}
\usepackage{multirow}

\title[Evaluating hydrodynamical simulations]
{Evaluating hydrodynamical simulations with green valley galaxies}
\author[Angthopo et al.]
{J. Angthopo$^{1}$\thanks{E-mail: james.angthopo.16@ucl.ac.uk}, A. Negri$^{2,3}$,
I. Ferreras$^{2,3,4}$, I.~G. de la Rosa$^{2,3}$, \and
C. Dalla Vecchia$^{2,3}$, A. Pillepich$^5$
\\
$^1$ Mullard Space Science Laboratory, University College London, 
Holmbury St Mary, Dorking, Surrey RH5 6NT, UK\\
$^2$ Instituto de Astrof{\'i}sica de Canarias, Calle V{\'i}a L{\'a}ctea s/n,
E38205, La Laguna, Tenerife, Spain\\
$^3$ Departamento de Astrof{\'i}sica, Universidad de La Laguna (ULL),
E-38206 La Laguna, Tenerife, Spain\\
$^4$ Department of Physics and Astronomy, University College London,
Gower Street, London WC1E 6BT, UK\\
$^5$ Max-Planck-Institut f\"ur Astronomie, K\"onigstuhl 17, D-69117 Heidelberg, Germany
}

\begin{document}
\date{Revised version. MNRAS ref. MN-20-3387-MJ. October 7, 2020.}
\pagerange{\pageref{firstpage}--\pageref{lastpage}} \pubyear{2020}
\maketitle
\label{firstpage}


\begin{abstract}
We test cosmological hydrodynamical simulations of galaxy formation
regarding the properties of the Blue Cloud (BC), Green Valley (GV) and
Red Sequence (RS), as measured on the 4000\AA\ break strength vs stellar
mass plane at $z=0.1$. We analyse the RefL0100N1504
run of EAGLE and the TNG100 run of IllustrisTNG project, by
comparing them with the Sloan Digital Sky Survey, while taking into
account selection bias.  Our analysis focuses on the GV, within
stellar mass $\log\,\mathrm{M_\star/M_{\odot}} \simeq 10-11$, selected
from the bimodal distribution of galaxies on the D$_n$(4000) vs stellar
mass plane, following Angthopo et al. methodology.  Both simulations match
the fraction of AGN in the green-valley. However, they over-produce
quiescent GV galaxies with respect to observations, with
IllustrisTNG yielding a higher fraction of quiescent GV galaxies than
EAGLE.  In both, GV galaxies have older luminosity-weighted ages with
respect to the SDSS, while a better match is found for mass-weighted
ages.  We find EAGLE GV galaxies quench their star formation early,
but undergo later episodes of star formation, matching
observations. In contrast, IllustrisTNG GV galaxies have a more
extended SFH, and quench more effectively at later cosmic times,
producing the excess of quenched galaxies in GV compared with SDSS, based on the 4000\AA\ break
strength. These results suggest the AGN feedback subgrid physics, more
specifically, the threshold halo mass for black hole input and the
black hole seed mass, could be the primary cause of the
over-production of quiescent galaxies found with respect to the
observational constraints.
\end{abstract} 


\section{Introduction}
\label{Sec:Intro}
Galaxy formation and evolution represents one of the key important
frontiers of astrophysics over the past decade.  Owing to the
complexity of physics concerning the transformation of gas into stars,
a number of open questions remain. In order to advance the field, a
combination of two main approaches are essential: (i) high quality
surveys, most notably SDSS \citep{SDSS, 2006Gunn}, and (ii)
cosmological hydrodynamical simulations, such as
EAGLE \citep{Schaye:15} or IllustrisTNG
\citep{Pill:2018, Mari:2018, Springel:18, Nel:18, Jill:2018}.
Both methods complement each other, as observations help to constrain
various parameters in simulations to reproduce the fundamental
properties of galaxies, while simulations enable the physical
interpretation of the observations.

Large surveys, such as SDSS, that combine photometry and spectroscopy, enabled
the discovery of fundamental properties of galaxies i.e. their bimodal
distribution \citep{Strateva:01},  in colour-magnitude
\citep{GravesBimod, Martin:07},
star formation rate (SFR) - mass \citep{Schim:07}, UVJ bi-colour
\citep{2009Will} and colour-mass \citep{Schawinski:14} diagrams. This bimodality
is thought to be due to the existence of two distinct types of
galaxies: Star Forming (SF) and Quiescent (Q). SF galaxies appear blue
in optical colours, and have substantial amounts of gas and dust.  In
contrast, Q galaxies are redder in the same colours, and feature low
amounts of dust or cold gas, resulting in little or no star formation
activity. Due to these differences, SF and Q galaxies occupy different
regions on the colour-mass plane. The region dominated by SF and Q
galaxies were coined Blue Cloud (BC) and Red Sequence (RS),
respectively.

The region situated between this bimodal distribution is thought to
represent a transition phase, termed the Green Valley \citep[hereafter
GV,][]{Salim:14}, where SF galaxies -- typically found in the BC --
slowly shift towards quiescence, gradually approaching the RS.
\cite{Faber:07} proposed several evolutionary paths to explain this
transition, mostly involving quenching of their star formation
activity. Moreover, galaxies are also able to obtain a fresh supply of
gas, via accretion or mergers, initiating subsequent episodes of star
formation (rejuvenation), that cause galaxies in the RS to move
``back'' into the GV/BC \citep{2005Th, Rejuv2010}, thus complicating
the study of galaxy evolution.

Owing to the sparsity of the GV, it is thought that galaxies traverse
rapidly this region \citep{Salim:14, Bremer:18}, therefore having short
quenching timescales. Observational constraints suggest a transition timescale of
$\tau_{\rm GV}$$\sim$100\,Myr and $\tau_{\rm GV}$$\sim$7\,Gyr for elliptical and
spiral galaxies, respectively, showing at least two modes of
quenching with a strong dependence on morphology \citep{Schawinski:14}.
A lower resolution study of GV galaxies, irrespective of
morphology, and adopting exponentially decaying star formation histories (SFH),
yields transition times in the range
$\tau_{\rm GV} \sim$2--4\,Gyr \citep{Phill:19}.
Similar results are obtained
when the GV is defined using the spectroscopic index D$_n$(4000)
\citep[][hereafter A19 and A20]{Angthopo:19, Angthopo:20}, finding
transition times $\tau_{\rm GV}$$\sim$1.0--3.5\,Gyr, for Q galaxies, and
$\tau_{\rm GV}$$\sim$0.5--5.0\,Gyr, for all types of galaxies.
Furthermore the use of specific star formation rate (sSFR) or
4000\AA\ break strength to define the GV have hinted at shorter timescales when
evolving from BC to GV with respect to the GV to RS
transition \citep[][A20]{Salim:14}.
Alternatively, the use of sub-millimeter
fluxes to explore galaxy evolution, rather than optical light,
suggests the presence of an overdensity in lieu of a valley, termed
the  ``green mountain'', which
questions the concept of rapid quenching \citep{Eales:18}.
Note that by using longer wavelengths, the analysis shifts towards dust
emission, and therefore focuses on the ``active'' sample. In optical light
we probe more directly the contrast between the ``active'' and the quenched
phases of evolution.

From a theoretical standpoint, state-of-the-art hydrodynamical
simulations such as EAGLE \citep{Schaye:15} and IllustrisTNG
\citep{Pillepich:18} are able to 
reproduce the general
fundamental properties of galaxies i.e. the evolution of the galaxy
mass function \citep{Fur:15,Kav:17,Pillepich:18}, AGN
luminosity  \citep{RG:2016,Vol:16,Mc:2017}
as well as the bimodality of galaxy
colour \citep{Tray:15,Tray:16,Nel:18}, and the SFR and UVJ-based
quenched fraction at $z\lesssim2-3$ \citep{Don:19,Don:20a,Don:20b}.
Studies based on simulations estimate a transition timescale ($\tau_{\rm GV}$)
that depends on stellar mass, similar to those derived from
observational constraints,
with $\tau_{\rm GV}$ $\sim$3\,Gyr at low stellar mass
($\log$\,M/M$_{\odot}<$9.6).  Intermediate stellar mass galaxies
($9.7\lesssim\log$\,M/M$_{\odot}$ $\lesssim$10.3) have longer
transition timescales, whereas the most massive galaxies
($\log$\,M/M$_\odot\gtrsim$10.3) yield the lowest transition
timescales $\tau_{\rm GV}\lesssim$2\,Gyr \citep{Wright:18}. More
detailed studies, reaching a higher mass resolution and segregating with
respect to morphology \citep{Tach:19, Correa:19}, suggest elliptical
galaxies have slightly lower quenching timescales $\tau_{\rm GV}$
$\sim$1.0\,Gyr, in comparison to disc-type galaxies $\tau_{\rm GV}$
$\sim$1.5\,Gyr.

Both observations and simulations show at least two evolutionary
channels, as elliptical galaxies seem to quench rapidly, while spiral
galaxies have a gradual decrease of the SFH, where they slowly exhaust
their gas
supply \citep{GalZooSme2015}. Furthermore, velocity dispersion (or
stellar mass) and galaxy structure, i.e, 
concentration, central density 
and effective density, seem to be the fundamental properties 
associated with galaxy evolution 
\citep{Gallazzi:05, 2009Gr, Star:2019, Barro:2017, ChenFab:2020}, 
thus the quenching
timescales and the physical mechanism behind quenching of star
formation are heavily dependent on velocity dispersion 
and galaxy structure. However the primary mechanism for quenching star 
formation remains a heavily debated topic. From the theoretical 
perspective, at low to intermediate stellar mass 
($\log$\,M/M$_{\odot}$ $\lesssim$10.5)
supernova-driven feedback \citep{DS:86,DV:2012}, radio-mode AGN
\citep{2006LateAGN} and environmental effects such as ram pressure 
stripping strangulation and harassment \citep[see, e.g.][]{AP:15} 
seem to be the dominant form of quenching. In contrast, 
massive galaxies ($\log$\,M/M$_{\odot}\gtrsim$10.5), 
appear to quench star formation with a
combination of quasar-mode AGN \citep{Schawinski:07,Dashyan:19,Man:19}
and major mergers \citep{2006EarlyAGN}. Furthermore, galaxies with
halo mass above a critical value 
M$_{\rm crit} \sim 10^{12}$M$_{\odot}$,
halo quenching is also thought to play an
important role \citep{Faber:07,Daniel2014}. Galaxy morphology plays
an important role in the primary physical mechanism for quenching: for
elliptical galaxies, AGN feedback is essential, while disc galaxies
are sensitive to other quenching mechanism
\citep{Dashyan:19, Correa:19}.

Simulations published over the last 5-6 years have succeeded in
reproducing the general properties of galaxies.  However even more
detailed analyses reveal a number of mismatches between simulations 
and observations, and provide a way to advance our knowledge in such a
complicated subject. Due to the complexity of simulations,
we focus our analysis specifically 
on stellar population properties and try to infer
how the subgrid physics affects galaxy formation and evolution. 
GV galaxies constitute a very informative subsample, where
the effects of subgrid physics can be tested, regarding the varying
stellar population content as galaxies evolve. Furthermore, since 
the GV is a transition region where the most fundamental quenching
mechanisms operate, this comparison allows us to explore the
prescriptions adopted by the modellers to trigger quenching.
Hence we compare the observational constraints with 
state-of-the-art-simulations (EAGLE and IllustrisTNG)
following our robust definition of GV, as presented in 
A19 and A20, based on the 4000\AA \ break strength.
The paper is laid out as follows: Section \ref{Sec:Sample} outlines 
the two simulations and the survey used for the analysis. Section
\ref{Sec:method} looks into pre-processing of the 
data to avoid selection biases. Section \ref{Sec:EGV} presents the
comparison of the simulated galaxies with the observational
constraints. Finally in sections \ref{Sec:DnC} and \ref{Sec:Conc} we
discuss the main results and summarise them.

\section{Sample}
\label{Sec:Sample}

We present here some details of the simulation and observational data
explored in this paper. We refer the reader to the relevant papers
quoted below for more details regarding the sample. We will focus
our description here on some of the aspects more relevant to the
analysis of green valley galaxies, especially on the way AGN feedback
has been implemented in the simulations.

\subsection{The EAGLE (RefL0100N1504) simulation}
EAGLE \citep{Schaye:15, Crain:15} represents a set of cosmological
hydrodynamical simulations comprised of multiple runs with different
box sizes and resolution. We use here the fiducial EAGLE simulation
RefL0100N1504 (hereafter Ref100), that adopts a comoving box size of
L=68$h^{-1}$\,Mpc=100\,Mpc, containing $1054^3$ dark matter (DM)
particles, with a baryonic particle mass of
$m_g$=1.81$\times10^{6}$M$_\odot$ and dark matter particle mass of
$m_{\rm dm}$=9.70$\times10^6$M$_\odot$.  EAGLE is
based on a modified version
of GADGET~3 \citep{Springel:05}, in terms of the implementation of
subgrid physics, the smoothed particle hydrodynamics (SPH)
formulation and the choice of time
steps. EAGLE adopts a flat $\Lambda$CDM cosmology, taking the
parameters derived from \citet{Planck:14}, just for reference,
$\Omega_m$=0.307, $\Omega_\Lambda$=0.693, $\Omega_b$=0.048,
$h$=0.6777, $\sigma_8$=0.8288.

Over galaxy scales, EAGLE, as all other
galaxy simulations, depend on a number of
prescriptions collectively termed ``subgrid physics'' that aim at
describing: radiative cooling and photoheating; reionization of
hydrogen; star formation; stellar mass loss and Type Ia Supernovae;
feedback due to star formation and AGN; and growth of supermassive
black holes (SMBH).  The EAGLE simulation follows \citet{Schaye:08},
where the star formation rate depends on the gas pressure rather than
its density, better reproducing the observed Kennicutt-Schmidt
law. Due to the lack of resolution, simulations suffer from an
'overcooling' problem when considering stellar feedback, ineffective
at forming the observed high mass galaxies. To compensate for this, a
method following \citet{Vecchia:12} is implemented, that makes stellar
feedback a stochastic process, thus enabling the control of energy
accessible per feedback event.  A black hole seed with mass $m_{\rm
BH}=10^5h^{-1}$M$_\odot$ is included in the simulated galaxies by
converting a gas particle with the highest density to a collisionless
particle. This is applied to any galaxy with a dark matter halo mass
above $m_{\rm h,thres}$=$10^{10}h^{-1}$M$_\odot$. AGN feedback is treated in a
similar manner to stellar feedback -- energy is injected
stochastically. However, note that it is generally understood that
there are two major modes of AGN feedback: 'quasar'- and 'radio-'
mode. Due to the lack of resolution of simulations at present, they
are unable to distinguish between the
two \citep{NaabOstr:17}. Therefore in EAGLE simulations, only one mode
of AGN is implemented.  The chosen method behaves like quasar-mode
feedback, because the input thermal energy rate is proportional to the
gas accretion rate at the location of the SMBH.

EAGLE adopts the Bondi mass accretion rate, defined in \cite{BH:44}, namely:
$\dot{m}_{\rm Bondi}=(4\pi G^2 m_{\rm BH}^2 \rho)/(c_s^2 + \nu^2)^{3/2}$,
where $\rho$ represents the density of the gas near the SMBH. The parameter
$\nu$ is the relative velocity of the gas with respect to the black hole, and
$c_s$ is the sound speed. This accretion rate is used to
calculate the mass accretion rate as follows:
\begin{equation}
        \dot{m}_{\rm accr} = \dot{m}_{\rm Bondi} \times \mathrm{min}
        (C_{\rm vis}^{-1}(c_s/V_{\phi})^3, 1),
        \label{eq:Massaccr}
\end{equation}
where $C_{\rm vis}$ is a free parameter that relates to the viscosity
and is set at $2\pi$ for this simulation; $V_{\phi}$ represents the
rotation of the gas around the black hole. Therefore, unlike
IllustrisTNG (see below), EAGLE takes into account the angular
momentum of the particles \citep{RG:15}. The accretion rate is
contrasted with the Eddington mass accretion, where the minimum of the
two values is chosen to describe black hole growth.  The Eddington
mass accretion rate is defined by:
\begin{equation}
        \dot{m}_{\rm Edd} = \frac{4\pi Gm_{\rm BH}m_{p}}{\epsilon_r \sigma_T c},
        \label{eq:MasEdd}
\end{equation}
where $\epsilon_r = 0.1$ is the radiative accretion efficiency, and
$\sigma_T$ is the standard Thomson cross-section. This can be 
converted to an AGN luminosity via
$L_{\rm AGN} = \dot{m}_{\rm BH} \epsilon_r c^2$,
and the Eddington Luminosity is
$L_{\rm Edd} = 1.25 \times 10^{38}(m_{\rm BH}/\mathrm{M_\odot})$.
More detail on how the various subgrid physical
mechanisms are implemented in EAGLE can be found in \cite{Schaye:15}
and \cite{Crain:15}.  Note the simulated data used to create the
photometric/spectroscopic
equivalents of the SDSS measurements are computed within 
the central 3\,kpc of a galaxy, that corresponds to the mapping on
the galaxies of the aperture size of the SDSS fibres. However, the
stellar mass of the galaxy, which is used as the major parameter
to characterize the overall properties of a galaxy, is determined
for the whole system.

\subsection{The IllustrisTNG (TNG100) simulation }

IllustrisTNG represents an improved simulation over the original
Illustris project \citep{Vogel:14, Gen:14}, based on the AREPO code
\citep{Springel:10}. For this project, we will make use of
the TNG100 run, with publicly available data
\citep{TNGPR:2019}, with a box size
L=75$h^{-1}$\,Mpc$\sim$110\,Mpc,
with equal number of inital gas cells and dark matter particles,
N$_{\rm gas}$=N$_{\rm DM}$=$1820^{3}$, baryonic mass 
$m_{\rm baryon}$=$1.4\times10^{6}$M$_\odot$ and dark matter mass 
$m_{\rm dm}$=$7.5\times10^{6}$M$_\odot$
\citep{Pill:2018, Mari:2018, Springel:18, Nel:18, Jill:2018}.
The initial conditions of the TNG100 simulation are set at redshift
$z=127$, with cosmological parameters  
$\Omega_m = \Omega_{\rm dm} + \Omega_b$=0.3089, $\Omega_b$=0.0486,
$\Omega_\Lambda$=0.6911, $h$=0.6774, $\sigma_8$=0.8159,
in accordance with constraints proposed by the Planck
collaboration \citep{Planck:16}.  Similarly to EAGLE, and the original
Illustris simulation, the subgrid physics consists of radiative
cooling, star formation and SN feedback, black hole formation and
growth along with AGN feedback. However in contrast to Illustris,
IllustrisTNG made key improvements on three areas - stellar evolution
and gas chemical enrichment, growth and feedback of supermassive BHs
and galactic winds \citep{Wein:18, Pillepich:18}.

Star formation in TNG100 proceeds through gas stochastically
converting to star particles if their gas density grows above a
critical threshold $n_H$=0.13\,cm$^{-3}$. The number is tuned so that
it reproduces the observed Kennicutt-Schmidt law. Each star particle
is treated as a single-age stellar population with a \citet{Chab:03} initial
mass function. These populations evolve with time, eventually returning
a fraction of their mass and elements to the surrounding interstellar medium.
TNG100 incorporates both types of supernovae: core collapse
SNII, between the mass range $m_{\star}$=8--100\,M$_\odot$,
as well as type~Ia SN. 
For stellar mass between 1--8\,M$_\odot$, stars are assumed to evolve 
through an AGB phase. More detailed information can be found in \citet{Pillepich:18}.

Regarding AGN feedback, when the galaxy is in the low-accretion state
TNG100 uses kinetic AGN
feedback rather than the bubble model \citep{Sijacki:07}.
At high accretion rates TNG100 uses, analogously to EAGLE,
a thermal feedback model with continuous energy injection.
The initial black hole seed mass is 
$8\times 10^{5}\,h^{-1}$M$_\odot$, and the halo mass threshold is 
M$_{\rm h,thres}$=$7.38\times 10^{10}$M$_\odot$. This difference with respect to
EAGLE may be an important one that we present in the discussion, as a
potential cause of the discrepancies found between these two simulations
on the green valley.
The accretion rate implemented
in TNG100 follows a similar prescription to EAGLE. However, 
TNG100 distinguishes between two modes of AGN feedback,
kinetic and thermal, by use of the Eddington ratio given by:
\begin{equation}
        \lambda_{\rm Edd} = \frac{\dot{m}_{\rm Bondi}}{\dot{m}_{\rm Edd}}.
        \label{eq:Illacc}
\end{equation}
The Bondi accretion rate, which is the same as the mass accretion rate
($\dot{m}_{\rm accr}$), is formulated as:
\begin{equation}
        \dot{m}_{\rm Bondi} = \frac{4\pi G^2 m_{\rm BH}^2 \rho}{c_s^2},
        \label{eq:BHbond}
\end{equation}
where for this simulation, $c_s$ encapsulates both thermal and magnetic
signal propagation, therefore
$c_s = (c^2_{\rm s,therm} + (B^2/4\pi\rho))^{1/2}$. The Eddington mass accretion rate
is slightly different to the one defined in EAGLE (equation~\ref{eq:MasEdd}), as the
radiative efficiency of a black hole is set at $\epsilon_r$=0.2 \citep{Pillepich:18},
whereas EAGLE adopt $\epsilon_r$=0.1 \citep{Crain:15}. The AGN feedback mode
is set by choosing a threshold $\chi$, so that $\lambda_{\rm Edd}<\chi$
will result in kinetic feedback, whereas in the high accretion state,
$\lambda_{\rm Edd}\geq\chi$, thermal feedback will be
enforced \citep{Weinberger:17}. The value of the threshold,
$\chi$, is given by:
\begin{equation}
        \chi = \mathrm{\rm min}\left[\chi_0
        \left(\frac{m_{\rm BH}}{10^{8} M_{\odot}} \right)^{\beta}, 0.1 \right],
        \label{eq:Illchi}
\end{equation}
thus introducing a mass dependency. Both $\chi_0$ and $\beta$ are
free parameters. The limit of 0.1, shown in the equation above, allows
any black hole, regardless of mass to have a high
accretion rate \citep{Wein:18}.
Similarly to EAGLE, all parameters, including
the spectroscopic index D$_n$(4000), are measured within an aperture
corresponding to the central 3\,kpc of a galaxy, except for (total)
stellar mass, which is measured over the whole galaxy.

\subsection{Observational data (Sloan Digital Sky Survey)}
\label{sec:SDSS}

Following from A19 and A20, our observational constraints on the green
valley rely on the classic spectroscopic database of the
Sloan Digital Sky Survey DR14 \citep[][hereafter SDSS]{2006Gunn,DR14}.  This catalogue
consists of a subsample of galaxies with $r$-band magnitude between
14.5 and 17.7\,AB, selected for spectroscopic follow-up from the SDSS
photometric data.  The spectra cover the 3,800--9,200\AA\ window,
at resolution $R\equiv\lambda/\Delta\lambda$ (FWHM) of 1,500 at
3,800\AA, and 2,500 at 9,200\AA\ \citep{Smee:13}.
To minimise any bias associated with
redshift, we restrict the sample to $0.05<z<0.1$. We refer the
reader to A20 for details regarding aperture effects and other
potential biases of the selected sample.  Finally to get reliable
measurements of the spectral features, we impose a threshold in
signal-to-noise ratio, snMedian$\_{\rm r} >$10. Those constraints
result in a total set comprising $\sim$228,000 spectra.

In A19 and A20, we chose the velocity dispersion as the fundamental parameter
that correlates with the population properties, following the well-known
observational trend \citep[see, e.g.,][]{Bernardi:03,SAMIIg:19}. Velocity
dispersion is a ``clean'' observable in spectra with high S/N, with
significantly fewer uncertainties than stellar mass -- another important
parameter that correlates strongly with population properties. However,
in simulated data, the velocity dispersion of a galaxy is a complicated
quantity that depends on many details of the formation process, especially
the dynamics associated to the mass accretion history.  Stellar mass
provides a comparatively more robust indicator of the global
properties of a galaxy in a simulation. Therefore, in contrast with
A19 and A20, we adopt stellar mass as the main parameter in the
analysis of the trends. We convert the six velocity dispersion bins
of our previous study, 70$<$$\sigma$$<$250\,km\,s$^{-1}$, to stellar
mass bins, by use of the observed trend
$\log\,$M$_\star$/M$_\odot$=$(1.84\pm0.03)\log\sigma_{100}+(10.3\pm0.3)$,
derived from stellar masses as quoted in the SDSS galSpecExtra catalogue
\citep{Jarle:04}, where $\sigma_{100}$ is the velocity dispersion,
measured in units of 100\,km\,s$^{-1}$. Tab.~\ref{tab:SigtologM} shows
the mass bins derived using this conversion, along with
the newly defined blue cloud (BC), green valley (GV) and red sequence (RS).
Note this conversion between velocity dispersion
and stellar mass is solely done for the stellar mass bins. For
individual galaxies we use the stellar masses given by the
galSpecExtra catalogue \citep{Jarle:04}.

To define BC, GV and RS, we follow a 
data-driven approach, whereby at fixed stellar mass, we use 
the observed distribution of SF and Q galaxies, identified using the BPT
diagram \citep{BPT:81}, to define a probability distribution
function (PDF) for the BC ($\cal{P_{BC}}$) and RS ($\cal{P_{RS}}$), respectively.
The PDFs are assumed to follow a Gaussian distribution. From these, we
define the PDF of GV galaxies ($\cal{P_{GV}}$) as another Gaussian peaking at the
value of D$_n$(4000) for which $\cal{P_{BC}}=\cal{P_{RS}}$, i.e.
both subsets are indistinguishable at this value of the 4000\AA\ break strength.
The GV is further split into upper (uGV),
middle (mGV), and lower (lGV) green valley subsets, where we define the
different regions by the terciles of the distribution within each
stellar mass bin (see A19 and A20 for full details). 
We find good agreement between the BC, GV and RS defined with respect to
stellar mass or velocity dispersion, within statistical uncertainties.
The only significant deviation was found at the highest mass bin: 
10.93$<$$\log\,$M$_\star$/M$_\odot$$<$11.03, where a difference is 
found in the definition of the GV at the level $\Delta D_n(4000)$$\sim$0.1\,dex.
At the highest mass bins, $\log\,$M$_\star$/M$_\odot\gtrsim$10.5,
the BC is sparsely populated, therefore the probability-based
methodology is less accurate.

\begin{table}
  \centering
  \caption{Definition of blue cloud (BC), green valley (GV) and red sequence (RS)
  galaxy populations based on the 4000\AA\ break strength, as proposed in A19. This version
  corresponds to bins in stellar mass (instead of velocity dispersion), for a
  more appropriate comparison with simulations.
  Columns 2, 3 and 4 provide the D$_n$(4000) values associated with these regions.}
  \label{tab:SigtologM}
    \begin{tabular}{l|c|c|c} 
    \hline
    $\log M_{\star}/M_{\odot}$ & BC & GV & RS \\
    \hline
    10.03 -- 10.30 & 1.20 $\pm$ 0.08 & 1.40 $\pm$ 0.07 & 1.67 $\pm$ 0.13 \\
    10.30 -- 10.51 & 1.25 $\pm$ 0.09 & 1.47 $\pm$ 0.06 & 1.72 $\pm$ 0.11 \\
    10.51 -- 10.68 & 1.28 $\pm$ 0.10 & 1.52 $\pm$ 0.05 & 1.75 $\pm$ 0.10 \\
    10.68 -- 10.81 & 1.30 $\pm$ 0.10 & 1.54 $\pm$ 0.05 & 1.77 $\pm$ 0.09 \\
    10.81 -- 10.93 & 1.32 $\pm$ 0.10 & 1.57 $\pm$ 0.04 & 1.78 $\pm$ 0.08 \\
    10.93 -- 11.03 & 1.34 $\pm$ 0.12 & 1.60 $\pm$ 0.04 & 1.79 $\pm$ 0.08 \\ 
    \hline
    \end{tabular}
\end{table}

\begin{figure*}
  \centering
  \includegraphics[width=80mm]{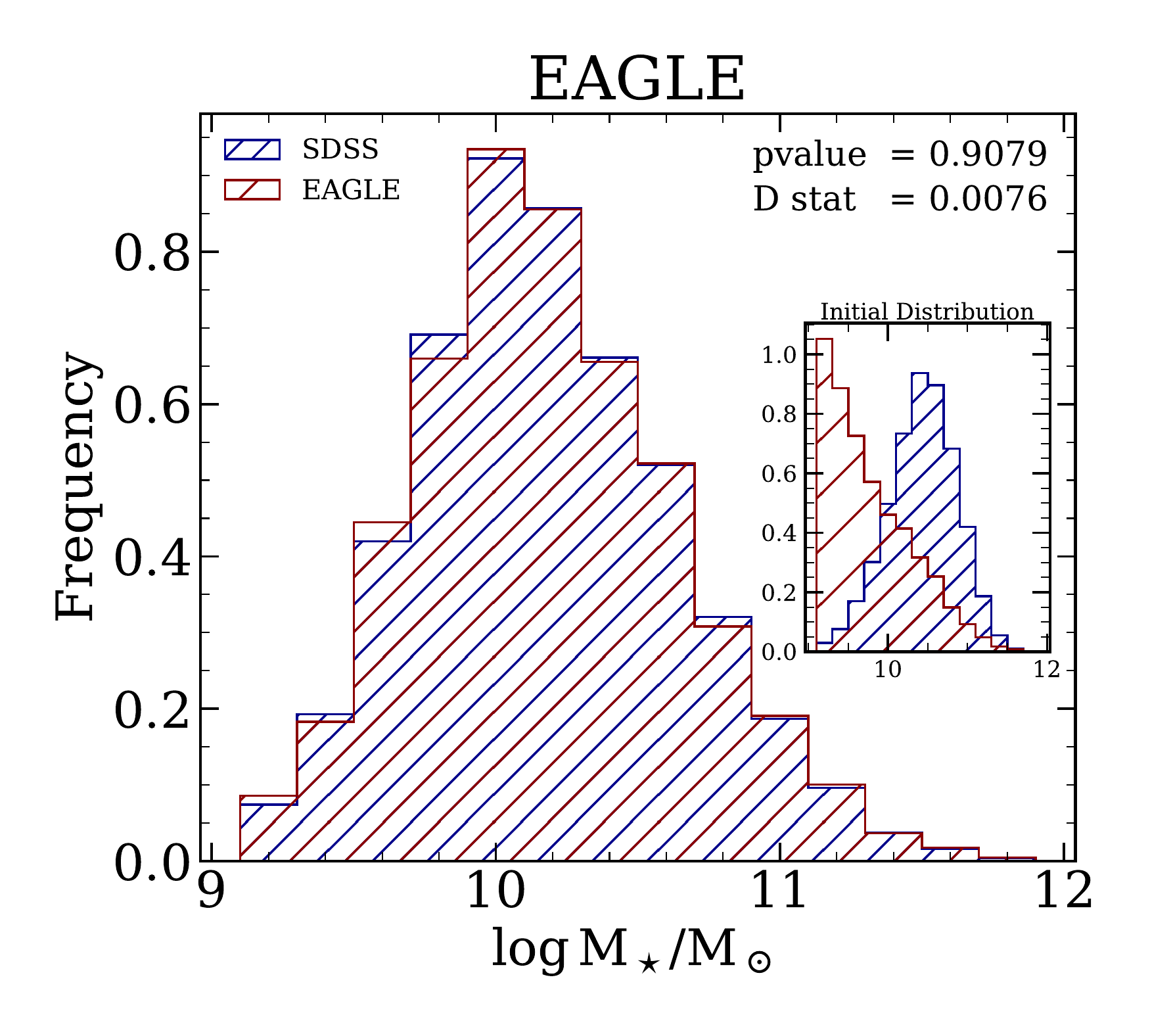}
  \includegraphics[width=80mm]{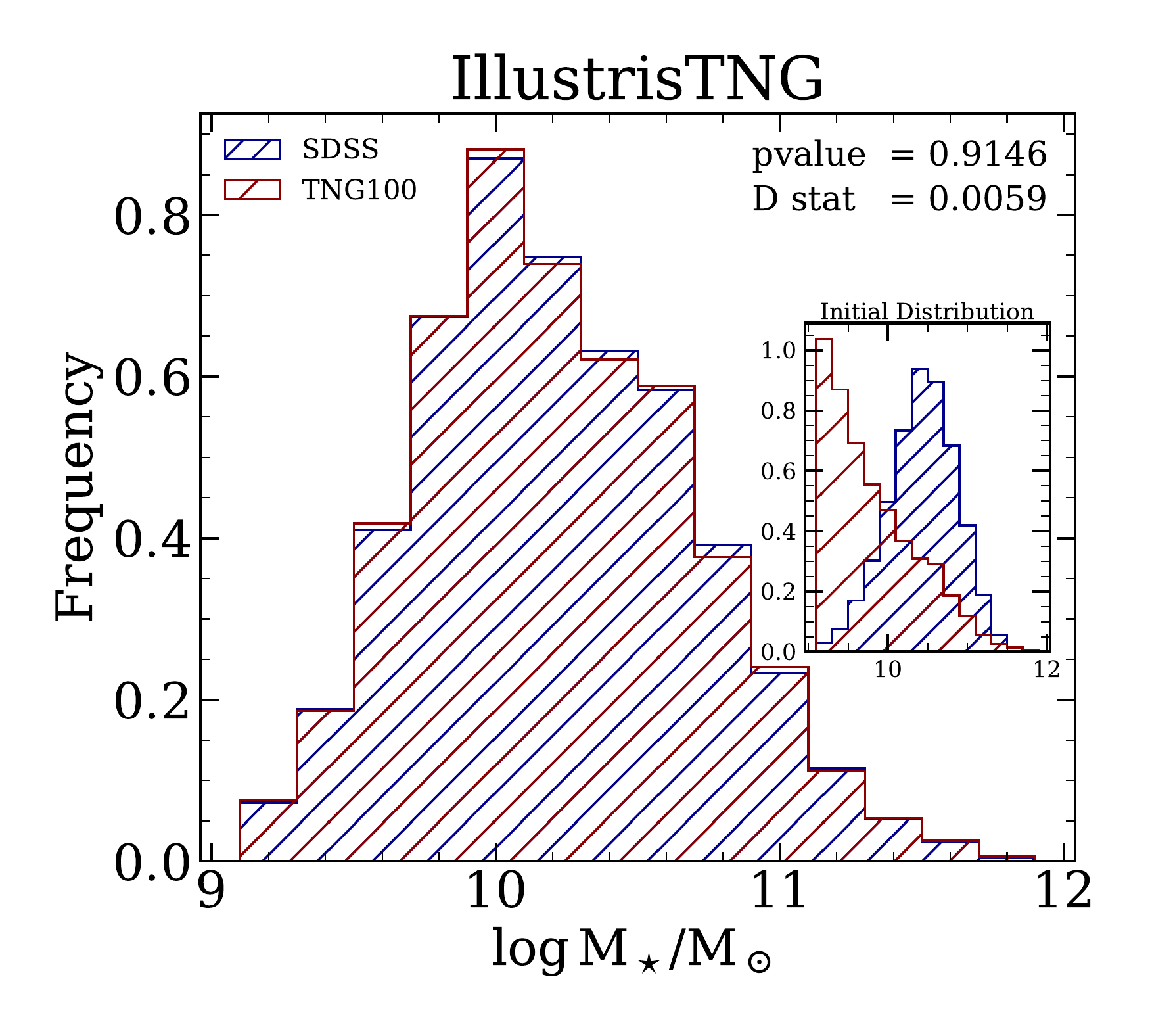}
  \caption{Distribution of 4000\AA\ break strength after
  homogenisation between observed (SDSS) and simulations data, showing
  EAGLE (\textit{left}) and IllustrisTNG (\textit{right}).  The blue
  and red histograms correspond to SDSS and simulations,
  respectively. The inset panels show the distribution of SDSS (blue)
  and simulation (red) galaxies before the homogenisation process.  A
  KS test confirms that these distributions originate from the same
  parent sample. Each panel shows the corresponding $D$ and $p$
  statistic of the test. Note that the SDSS sample requires a slightly
  different homogenisation sampling depending on the simulated set.}
\label{fig:Homg_Hist}  
\end{figure*}


\section{Data Pre-processing}
\label{Sec:method}
Before carrying out the analysis, we need to pre-process both the
observational and simulation data to homogenise their distributions
in order to minimise selection biases, and to produce synthetic
spectra from the simulated galaxies with the same instrumental
signature as the SDSS data. Note the spectra and the simulation
parameters are extracted within a R=3\,kpc aperture, to make a fair comparison
with SDSS, while the stellar masses, for both
simulation and SDSS, are derived for the whole galaxy. 
This section focuses on the methodology adopted to prepare the samples
so that a direct comparison could be made.

\subsection{Synthetic spectra}
\label{sec:SynthSpec}

Our comparison uses the $z=0.1$ snapshot of the simulations, comparable
to the redshift range of the SDSS database (differences between $z=0$
and $z=0.1$ are minimal for this analysis). In both EAGLE and TNG, we
produce a simple stellar population (i.e. single age and chemical
composition) for each stellar particle, mixing those populations into
composites for all particles in the same galaxy. The stellar particles
have mass above $10^6$M$_\odot$, therefore this approach is well
justified and does not suffer from any issues related to the sampling
of the IMF.  Due to the mass resolution of the stellar component in
state-of-the-art simulations at present, the low SFR regime is
typically sampled by the creation of a few or eventually just one
$10^6$M$_\odot$ star particles. Note these will in turn
produce a bias in the spectra, towards younger luminosity-weighted ages.

To produce spectra comparable to those from the 3\,arcsec fibre-fed
spectrograph of the classic SDSS data, we combine the spectra from all
stellar particles within a 3\,kpc galacto-centric
radius. For reference, the SDSS fibre radius maps a
physical distance of 2.8\,kpc at redshift $z=0.05$ for a vanilla
flavoured $\Lambda$CDM cosmology with $\Omega_m=0.3$, $h=0.7$.
The synthetic spectra are taken from the
population synthesis models E-MILES
\citep{eMILES}, based on a fully empirical stellar library
\citep{MILES} and
Padova isochrones \citep{Girardi:00}. The E-MILES spectra extend from
the far UV to the mid-IR (1680\AA\ to 5\,$\mu$m), spanning 
stellar ages from 6.3\,Myr to 17.8\,Gyr (we restrict the oldest ages
to the cosmological age of the Universe at the fiducial redshift, $\sim$12.4\,Gyr),
and metallicity ranging from [M/H]=$-$1.71 to $+$0.22. The models
adopt a \citet{Chab:03} IMF,  and we perform
a bilinear interpolation in age and metallicity of the E-MILES SEDs.
To avoid systematics caused by the modelling of dust attenuation,
and taking advantage of the insensitivity of the 4000\AA\ break strength
to typical amounts of dust in galaxies (see A20), we do not include
dust in the modelling of the synthetic spectra. Finally, the data are
convolved to the R$\sim$2,000 resolution of the SDSS classic spectrograph,
in quadrature with a Gaussian function mimicking the kinematic kernel
that corresponds to each stellar mass bin, following the trend
presented above, between stellar mass and velocity dispersion.

Previous work in the literature regarding the dust
modelling in EAGLE and TNG100 \citep{Tray:15, Tray:17, Nel:18} find a
level of agreement with the observational constraints. We emphasize in
this paper that the effect of dust on the spectra of galaxies represents
an additional layer of complexity in galaxy formation models that go
beyond the scope of this paper. Our main aim is to explore the evolution
of the stellar population properties across the green valley to probe
the more fundamental aspect of how feedback is implemented in models
to shape the star formation history of galaxies. The use of the D$_n$(4000)
index to define the location of the GV allows us to bypass the complexity
of dust, avoiding potential biases produced by the effect of dust attenuation
on other observables, most notably colours based on broadband filters.
Fig.~B1 of A20 shows a dust correction on the SDSS data causes a 
change in the position of the D$_n$(4000)-defined BC, GV and RS by less
than 0.06\,dex. Just as a test of the actual effect of dust attenuation
on our data, we applied the dust model described in Negri et al. (in prep.),
which is a modification of the dust model of \cite{Tray:15} for the EAGLE
simulation. We found negligible changes in the definition of the BC, GV, RS
when using the D$_n$(4000) index, as expected.

\subsection{Homogenisation of simulation and observation data}
\label{sec:homogenisation}

A fundamental step in the comparison between observations and simulations
involves ensuring that similar galaxy samples are considered.
Different selection effects in observations and simulations will yield
samples with incompatible distributions of stellar mass, hence the
need for a homogenisation process.  More specifically, the Malmquist bias imposed by the
r$<$17.77\,AB limit for spectroscopic follow-up \citep[see,
e.g.,][]{DR14} implies that low mass galaxies
(M$_\star\lesssim10^9$M$_\odot$) are missed in SDSS, with a
clear trend with redshift. In contrast, simulations are biased against
high mass galaxies (M$_\star\gtrsim 10^{10}$M$_\odot$) due to their
volume limitation \citep[see, e.g.,][]{Schaye:15}. Therefore for a
fair comparison between these data sets, we must ensure the
distributions are statistically compatible.

From the original samples, we select sets that are
``homogeneous'' regarding stellar mass -- which is the parameter we
assume to act as the major driver of the stellar population
content. Note we have to proceed with two different sets of
comparisons: one between SDSS and EAGLE and another one between SDSS
and TNG100. Moreover, we homogenise a pair of samples by finding
a \textit{pivot} stellar mass, so that below (above) this mass we
randomly exclude simulated (observational) galaxies, as we have a lower
fraction of observational (simulated) galaxies.  Here
the \textit{pivot} mass bin is chosen as the one for which there is a
greater fraction of observed galaxies in a mass bin compared to the
fraction of simulated galaxies. Fig. \ref{fig:Homg_Hist} shows the
histogram of homogenised mass distribution of EAGLE (\textit{left})
and TNG100 (\textit{right}) galaxies, with respect to SDSS. The blue and red
hatched areas represent SDSS and simulation histograms,
respectively.

To numerically assess the level of homogeneity between the respective
stellar mass distributions of observations and simulations, we carried
out a Kolmogorov-Smirnov test \citep[hereafter KS-test, see,
e.g.,][]{KStest}. When comparing the stellar mass distribution of the
original samples, we get high values of the $D$-statistic ($\sim$0.56)
between simulated and SDSS data (in both the EAGLE and TNG100
simulations), leading to a low probability that the samples are
produced by the same parent distribution. However, after
homogenisation, the $D$-statistic becomes, by construction, low
($\sim$6--7$\times 10^{-3}$) with high values of the probability
($\gtrsim$90\%) that the samples originate from the same
distribution. However, note that we compare SDSS individually with
either EAGLE or TNG100. We do not aim at creating a joint
SDSS-EAGLE-TNG100 homogenised sample as this will reduce further the
size of the working sample, restricting the stellar mass range.
Therefore we have two sets of SDSS galaxies: one for EAGLE and a
different set for TNG100.
We note this will lead to some differences between the two
sets of galaxy spectra from SDSS, quantified in Appendix \ref{sec:Frac_Match}.
Additionally, variations between the two simulation data sets are also expected
due to the different cosmological volumes probed, as well as the
prescriptions to model the subgrid physics that controls the stellar
mass growth in galaxies.  We emphasize that this procedure is needed
for a fully consistent comparison of the simulations with
observational data.  Note that in all samples, we make no distinction
between satellite and central galaxies, as an environment-related
analysis will be published in a separate paper (Angthopo et al., in
prep).

\begin{table*}
  \centering \caption{Criteria adopted to define galaxy activity: Seyfert/LINER
  AGN, quiescence (Q) or star-formation (SF). We constrain two
  parameters: the Eddington ratio ($\lambda_{\rm Edd}$) and the
  specific star formation rate (sSFR, defined as the ratio between
  the instantaneous star formation rate and the stellar mass).
  Here $x$ represents the parameter at the top of each
  column. Also shown are the percentage of each type of galaxies
  produced in the simulation that matches the selection criteria.
  In columns~4 and 7 we show, in the same column,
  both the simulated and SDSS fractions, the latter in brackets.}
  \label{tab:AGNdef}
    \begin{tabular}{l|ccc|ccc|} 
    \hline
    \multicolumn {1}{c}{} & \multicolumn{3}{|c|}{EAGLE} &  \multicolumn{3}{|c|}{TNG100} \\
    \hline
    Type & $\log(\lambda_{\rm Edd})$ & $\log$(sSFR/yr$^{-1}$)  & $\%$ & $\log(\lambda_{\rm Edd})$ &
    $\log$(sSFR/yr$^{-1}$) & $\%$ \\
    \hline
    Seyfert & $x \geq-2.0 $ & - & 3.86 (2.77) & $x \geq -1.4$ & -  & 3.03 (3.04) \\
    \hline
    LINER   & $-4.2\leq x < -2.0 $ & $x\leq -$ 10.5 &  9.33 (9.68) & $-4.2$ $\leq x \leq -1.4$
    & $x\leq -$ 11.0 & 8.69 (10.31) \\
    \hline
    \multirow{2}{*}{SF} & \multirow{1}{*}{$-4.2 \leq x <-$2.0} & \multirow{1}{*}{$x>-$10.5} & \multirow{2}{*}{62.99(62.91)} & \multirow{1}{*}{$-4.2 \leq x <-1.4$} & \multirow{1}{*}{$x > -11.0$} &
    \multirow{2}{*}{58.43 (60.69)} \\
    & $x<-4.2$ & $x>-11.0$ & & $x<-4.2$ & $x>-11.0$ & \\
\hline
    Q       & $x < -4.2$ & $x\leq -11.0 $ &  23.82 (24.64) & $x < -4.2$
    & $x \leq$ -11.0  & 29.85(25.97) \\
    \hline
    \end{tabular}
\end{table*}

\subsection{Galaxy Classification}
\label{sec:NebClassy}

Observationally, galaxies are traditionally classified regarding
nebular emission into quiescent (Q), star-forming (SF), Seyfert AGN,
LINER AGN, or composites (i.e. a mixture of those). Ratios of targeted
emission line luminosities allow us to separate the ionization environments
expected in the ISM of galaxies \citep{BPT:81, Fern:2011}. This is an
important classification scheme, as both star formation activity and
AGN are essential mechanisms in galaxy evolution, and the path to
quiescence still remains an open question
\citep[][A19]{Martin:07, Dashyan:19, Man:19}.
Our simple methodology does not take into account nebular emission when
creating the synthetic spectra from simulations. Therefore, we need to
get back to physical parameters in order to
classify the simulated galaxies as Q, SF or AGN. Regarding SMBH activity,
we compare the black hole accretion rate with respect to the Eddington ratio:
\begin{equation}
        \lambda_{\rm Edd} = \frac{\dot{m}_{\rm accr}}{\dot{m}_{\rm Edd}}.
        \label{eq:Edd_r_Gen}
\end{equation}
However, a non-trivial issue is to define the actual values of the
threshold ratio to segregate galaxies according to AGN activity.

Observational studies conclude Seyfert AGN have Eddington ratios
between $-2<\log\, (\lambda_{\rm Edd})
<-1$ \citep{Heck:04,Schulze:15,Geo:17,Ciot:17}, whereas lower
accretion rates correspond to radiatively ineffective AGN, i.e. LINER
or even no AGN. Some studies suggest a lower limit for LINER AGN
around $\log (\lambda_{\rm Edd}) \sim -6$ \citep{Heck:04, Li:17},
while others choose values as low as
$\log (\lambda_{\rm Edd}) \sim -9$ \citep{Ho:08, Ho:09}. Studies using
SDSS galaxies have calculated the Eddington ratio from [O{\sc III}]
emission, finding $\log (\lambda_{\rm Edd}) \sim -4$ \citep{KAGN:06}.
For the simulated galaxies, we adopt our own Eddington ratio limits,
along with specific star formation rate (sSFR), 
to classify galaxies as
Seyfert AGN, LINER AGN, SF (including composite) or Q.
Note the sSFR used here has been calculated
using the instantaneous SFR. An alternative selection would 
use the average SFR over some timescale, but previous work from the literature 
have shown that measuring SFR in different ways makes little
difference at low redshift \citep{Don:19, Don:20a}. Furthermore, 
the criterion to
select the corresponding values of $\lambda_{\rm Edd}$ and sSFR
is to impose equivalent {\sl global} ratios of Seyfert, LINER,
SF and Q to those found in the full homogenised SDSS samples. Note we
are interested in the {\sl relative} mass-dependent variation of these
fractions, and that the homogenisation in mass makes the comparison
between SDSS and the respective simulation meaningful.

EAGLE only imposes one form of AGN feedback, therefore it is not
possible to differentiate between Seyfert or LINER AGN activity.
From the simulations, we retrieve the black hole mass and their
accretion rate -- at the fiducial redshift $z=0.1$ -- and we use
Eq. \ref{eq:Edd_r_Gen} to find $\lambda_{\rm Edd}$ in each galaxy.
Comparing the total fraction of Seyfert AGN with respect to SDSS,
we obtain a threshold $\log(\lambda_{\rm Edd}) \gtrsim -2.0$.
We proceed similarly with the LINER galaxies, obtaining a range 
$-4.2 \lesssim \log(\lambda_{\rm Edd}) < -2.0$. Moreover, for 
LINERs we also impose a constraint on 
$\log$\,sSFR (yr$^{-1}$)$\lesssim -10.5$
to make sure SF activity is minimal. Note that most of the 
SDSS galaxies hosting LINERs have a sSFR similar to Q.
Galaxies with $ -4.2 < \log(\lambda_{\rm Edd}) < -2.0$
but with $\log$\,sSFR$>- 10.5$
are classified SF. Finally, galaxies with
$\log(\lambda_{\rm Edd}) \lesssim-4.2$ and $\log$\,sSFR$<-11$
($>-11$)
are considered to be Q (SF). These criteria  are summarised in 
Tab.~\ref{tab:AGNdef}, where the constraints are shown for both
EAGLE and TNG100 simulations. In each case, the table shows the
limits 
on $\lambda_{\rm Edd}$ and sSFR as well as the fraction of galaxies of
a given type in the homogenised simulations sample. The values in
brackets show the percentage of different types of galaxies observed
in the SDSS homogenised sample, identified via the BPT
diagram \citep{BPT:81, Kewley:01, Kauff:03b}. Both observation and
simulation data contain SF and composite galaxies (classified with a
BPT parameter either 1 or 3 in the SDSS galSpecExtra
catalogue of \citealt{Jarle:04}), while
we exclude low S/N star forming galaxies (BPT parameter 2).

The TNG100 simulation adopts two modes of AGN feedback --
thermal and kinetic feedback -- determined by Eq. \ref{eq:Illacc}
and \ref{eq:Illchi}, as defined in \cite{Wein:18}. The
fiducial approach fixes $\beta=2$ and
$\chi_0 = 2\times 10^{-3}$,
following \citet{Habo:19}.  However with this choice of parameters we
get too many Seyfert AGN galaxies.  To solve this, we carry out a
similar procedure to EAGLE, adopting our own thresholds. 
Doing so may introduce a bias, as we exclude Seyfert
AGN galaxies, and either consider them to be LINER AGN
or SF galaxies in the fraction estimates. However, note the
exclusion of Seyfert AGNs in the simulation data is justified,
as we find a low fraction of Seyfert AGN in the observations.
Furthermore, similarly to EAGLE, this is a necessary step
as we have to ensure the ratios of Seyfert, LINER, SF, and
Q galaxies are consistent, when comparing the properties of
observational and simulation data. The comparison between
the homogenised SDSS and TNG100 sets
gives $\log(\lambda_{\rm Edd}) \gtrsim -1.4$, slightly different to
that of EAGLE. This is due to different subgrid physics regarding
black hole growth and AGN feedback, as well as TNG100 having a
larger black hole seed. For LINER AGN we find
$-4.2 \lesssim \log(\lambda_{\rm Edd}) < -1.4$ gives consistent
fractions with respect to the observations. In TNG100, the
threshold $\log\,$sSFR$= -11.0$
gives a similar fraction of Q/SF galaxies with
respect to the observations. Note this choice 
produces a slightly higher percentage of Q galaxies in TNG100 (29.85$\%$) 
compared to the observations (25.97$\%$). Despite the slight 
variation in the selection criteria and in the homogenisation procedure, 
note the similarity of the parameter thresholds obtained to produce 
results matching the SDSS observations.

\begin{figure*}
  \centering
  \includegraphics[width=0.45\linewidth, height=75mm]{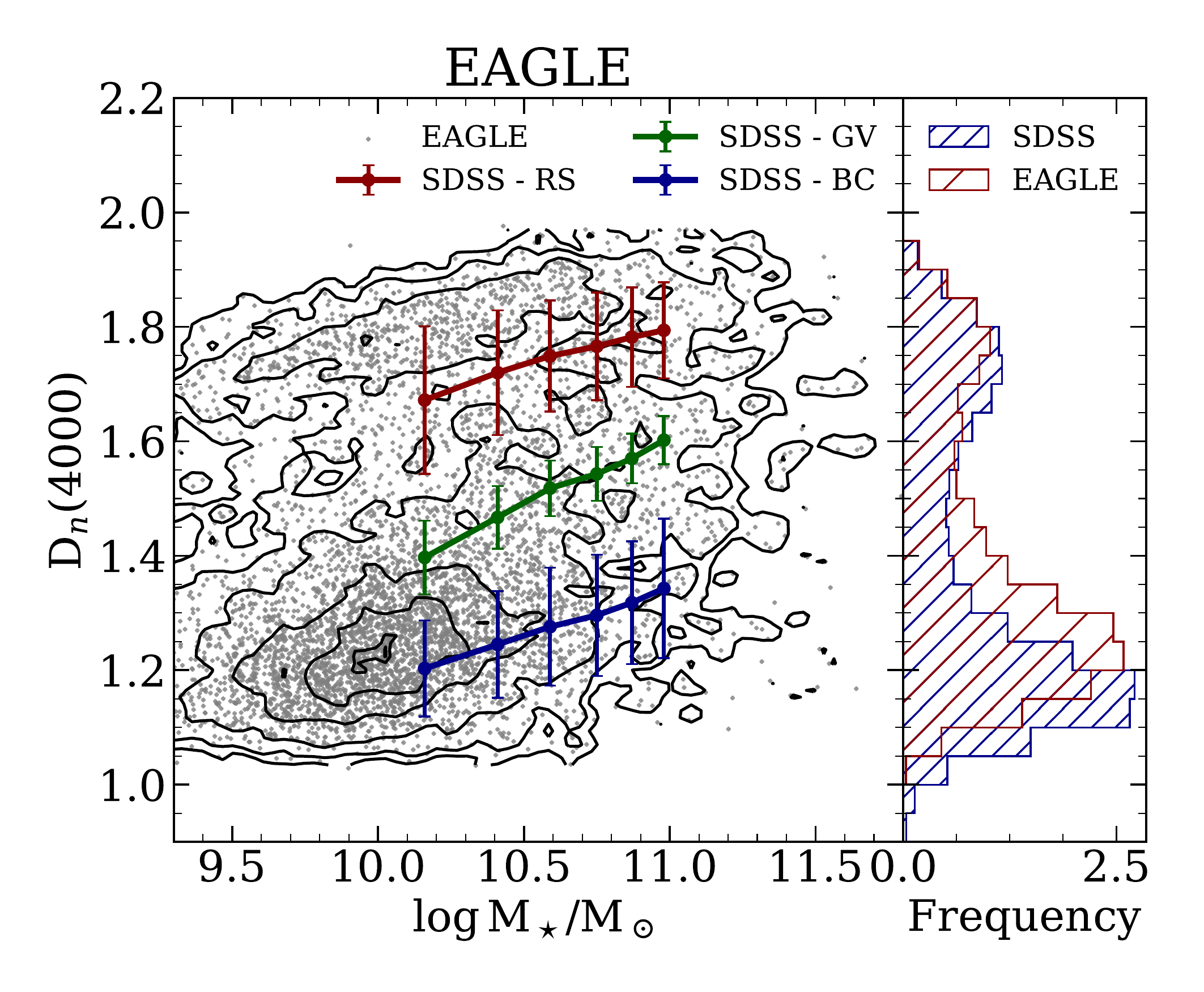} 
  \includegraphics[width=0.45\linewidth, height=75mm]{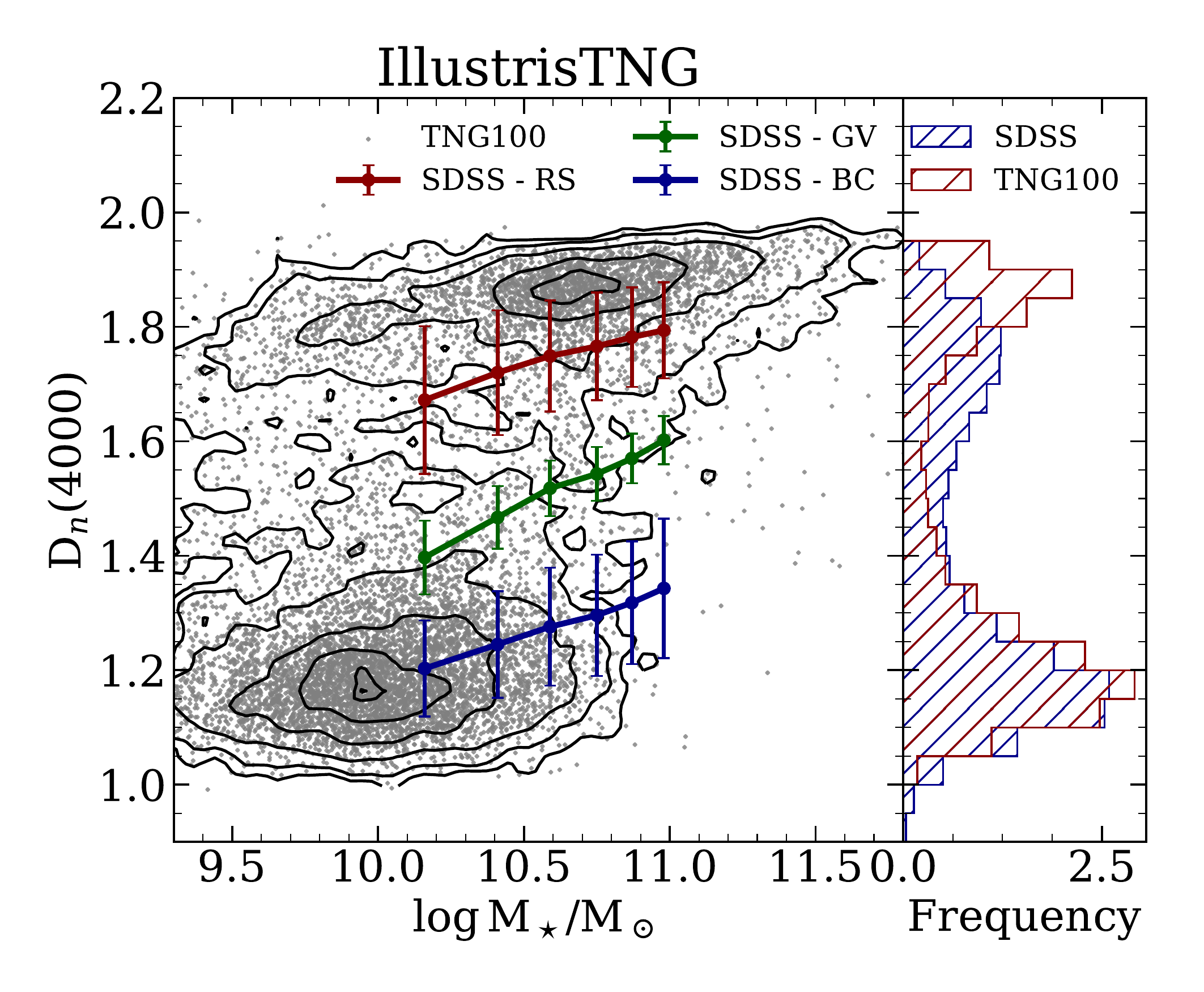}
  \caption{Distribution of galaxies on the D$_n$(4000) vs stellar mass plane.
  The gray dots represent the simulation data after homogenisation
  in EAGLE ({\sl left}) and IllustrisTNG ({\sl right}). The contours
  map areas of constant number density on the plot.
  The blue, green and red solid lines follow the observational definition
  of the BC, GV and RS, respectively. The error bars are given at one standard
  deviation from the mean. The histograms in the adjacent panels
  display the distribution of 4000\AA\ break strength in observations (blue)
  and simulations (red).}
  \label{fig:GV_D4k_Sim}  
\end{figure*}

\section{Confronting observations and simulations in the Green Valley}
\label{Sec:EGV}

The state-of-the-art simulations explored in this paper are capable of
matching the general fundamental properties of galaxies
\citep{Pillepich:18, Nel:18},
as well as the bimodality of galaxies in
colour \citep{Tray:15, Nel:18}.  We focus here on how well simulations
reproduce the bimodality of galaxies on the D$_n$(4000) vs
$\log$\,M$_\star$/M$_\odot$ plane. More specifically, we look at the
mass dependence of the fractions of AGN, SF and Q galaxies in the
GV. Given that the GV is a transitional region where quenching
processes efficiently drive galaxies to quiescence, this
comparison allows us to analyse the ability of the subgrid physics
imposed in the simulations to reproduce the observational data.

\subsection{Blue Cloud, Green Valley and Red Sequence}
\label{sec:BGR}

Our first analysis of the data -- before focusing on the GV --
involves a comparison of the BC, GV and RS between SDSS and
simulations. Note similar studies have been already performed on the
colour-mass plane, finding general agreement of the simulations with
observations \citep[see, e.g.,][]{Tray:16, Tray:17, Kav:17, Nel:18}
and on the SFR-mass plane \citep{Fur:15, Don:19}.
We revisit this comparison with our new definition of the BC, GV, RS 
regions (A19, A20), looking for hints that could help improve the
simulations.

\subsubsection{Comparison with EAGLE (RefL0100N1504)}
\label{sec:EAG_BGR}

Fig.~\ref{fig:GV_D4k_Sim}  shows the distribution of
EAGLE galaxies ({\it left}) on the D$_n$(4000) vs $\log$\,M$_\star$/M$_\odot$
plane, after sample homogenisation.
The blue, green and red data points
with error bars represent the observational (i.e. SDSS) BC, GV and RS.
The red and blue histograms on the side panels show the 
distribution of EAGLE and SDSS galaxies, respectively -- likewise for TNG100
on the rightmost panel. We find a mismatch in the distribution,
of about $0.1-0.2\,$dex at low D$_n$(4000),
and agreement at high D$_n$(4000).
Combining this with the stellar mass distribution, we see that
while there is an overall qualitative agreement, EAGLE 
seems to produce BC, GV and RS regions with higher D$_n$(4000),
and the mismatch increases from BC to RS.
This difference is greatest in the lowest mass bins,
$9.5\lesssim\log\,$M$_\star$/M$_\odot\lesssim$10.5.
The higher value of D$_n$(4000) at low-intermediate
mass could be due to the mass-metallicity relation, being shallower in EAGLE
with respect to the observations \citep{Schaye:15, Tray:17}.

One reason for the shallowness of the mass-metallicity relation could be due to
galaxies having a stronger chemical enrichment history at low mass
compared with the observations. D$_n$(4000) is sensitive to the metallicity
\citep{Balogh:99}, one of the manifestations of the age-metallicity degeneracy.
Moreover, the inclusion of only one mode of
AGN feedback --  resembling quasar-mode AGN, which happens to be dominant in high mass
galaxies  \citep{Shankar:06, Faber:07, GalZooSme2015} -- could offer
another potential explanation
for this trend. AGN feedback plays an important role in quenching of star
formation \citep{Martin:07, Goncalves:12, Wright:18, Dashyan:19},
therefore specially for lower mass galaxies, quasar mode AGN could
lead to quenching of star formation that operates too quickly. A
combination of these two effects could also cause the effects we see
here, where the red sequence features a shallower gradient with respect to
the observations.

\subsubsection{Comparison with IllustrisTNG (TNG100)}
\label{sec:ITNG_BGR}

In the TNG100 simulated galaxies (see Fig. \ref{fig:GV_D4k_Sim},
{\it right}), a better agreement is found in the BC with respect
to EAGLE; also evident from the histograms shown on the right.
However, there is significant disagreement in the GV and RS,
of  $\sim 0.1-0.3$\, dex.
Note the gradient of the RS on the D$_n$(4000) vs stellar mass plane
is shallower in TNG100 with respect to SDSS. It is also shallower that
the EAGLE data, with an excess of RS galaxies at low mass
($<10^{10}$M$_\odot$). Additionally, at the massive end (above
$10^{10.5}$M$_\odot$), a more drastic decrease in the number of BC
galaxies is apparent.  Note, the fraction of galaxies in the GV is
lower in TNG100 with respect to the observations,
specifically for $1.5 \lesssim$D$_n$(4000)$\lesssim 1.8$,
as shown by the histogram. Note that previous work in the
literature \citep{Nel:18} find better agreement when looking
at bimodality with respect to (g--r) colour.  However,
at lower stellar mass,
$9\lesssim \log\, \mathrm{M_{\star}/M_{\odot}} \lesssim 10$, they also find
a surplus of galaxies in the simulated red sequence, suggesting
over-quenching. More specifically, at low stellar masses, recent
comparisons find a shallower mass-metallicity trend with respect to
the observations \citep{Nel:18}, which might offer a possible explanation
to the excess of RS galaxies in TNG100.

\begin{figure*}
  \centering
  \includegraphics[width=82mm, height=85mm]{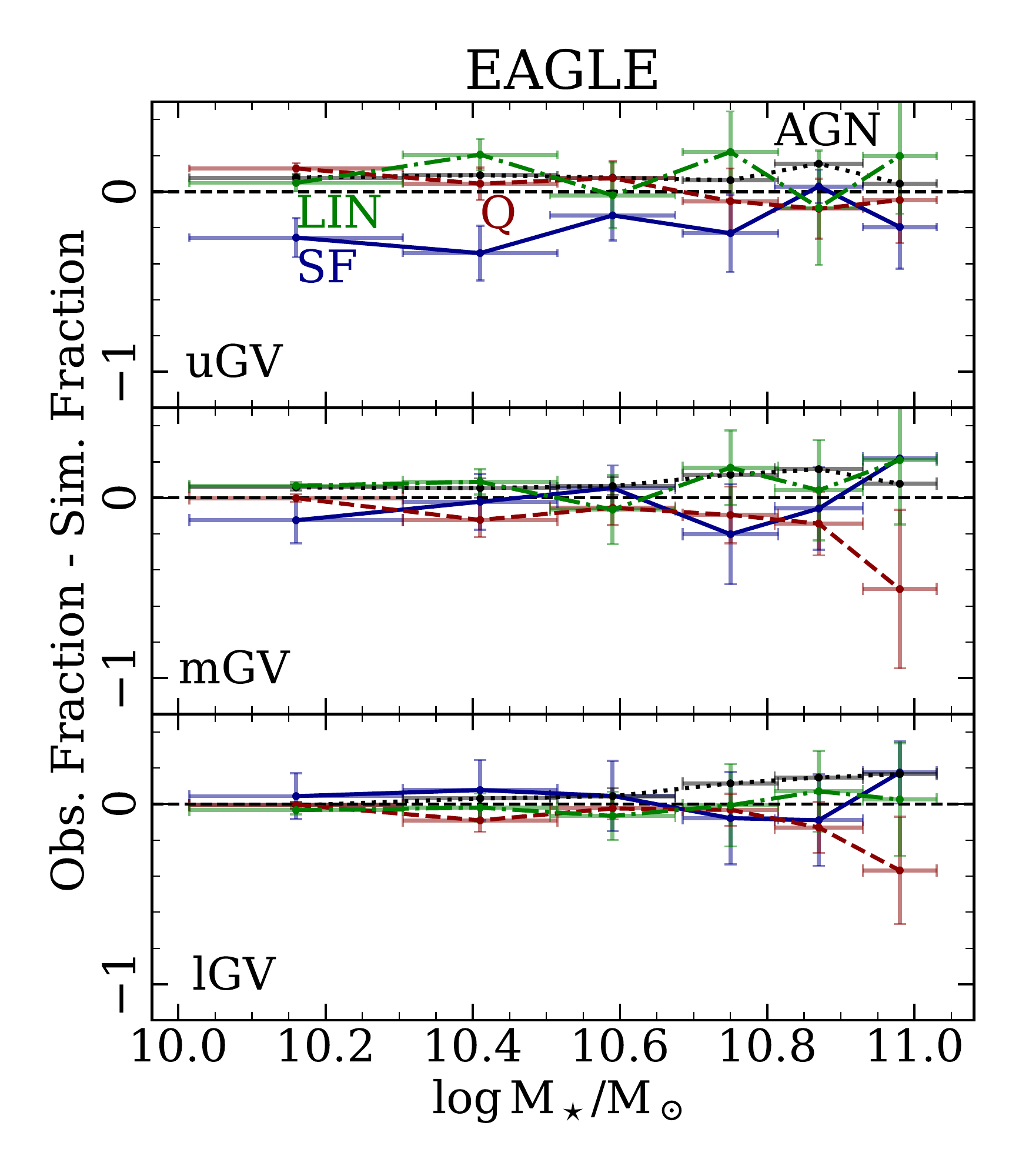}
  \includegraphics[width=82mm, height=85mm]{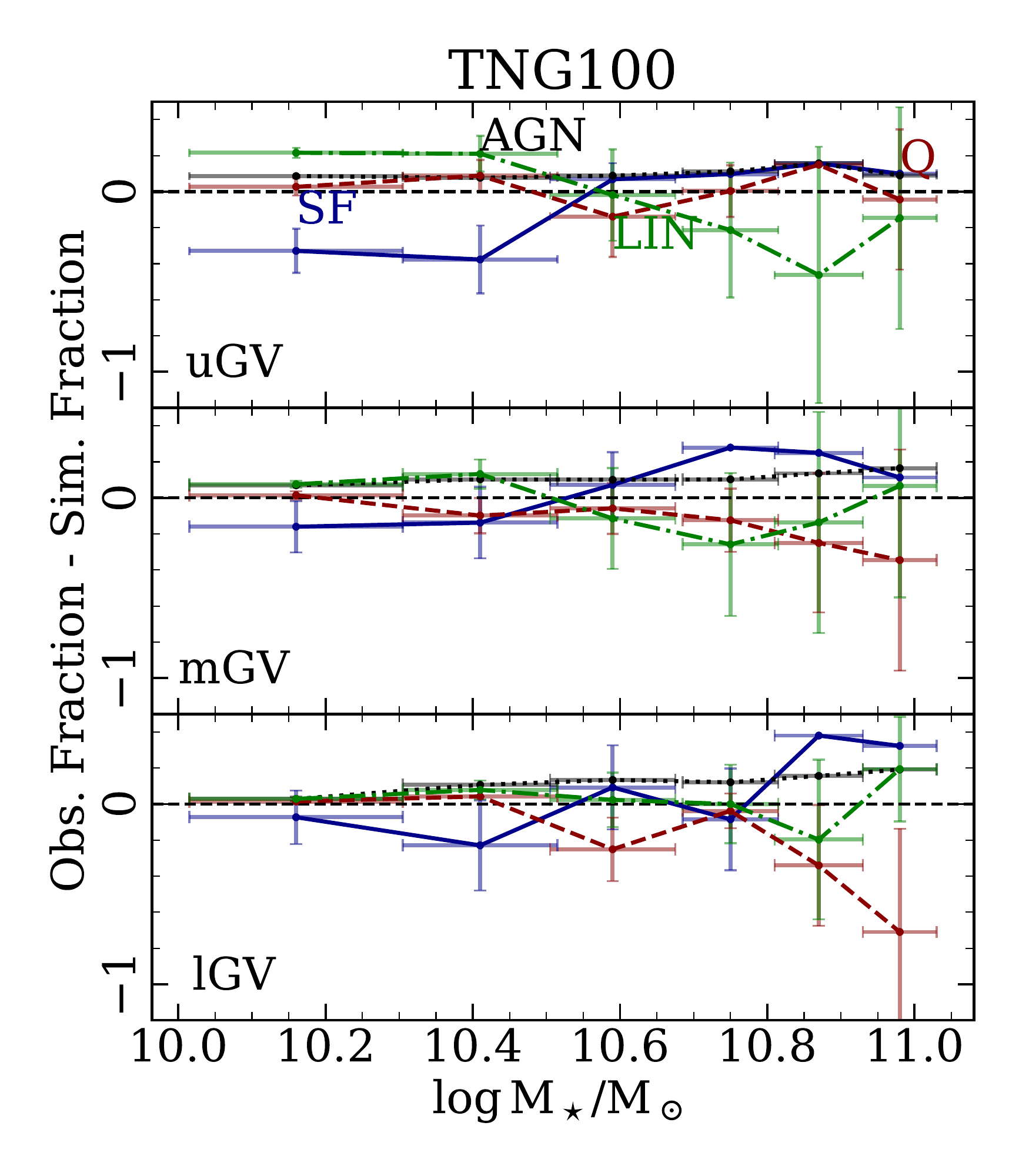}
  \caption{Fractional difference between observations and simulations
    of GV galaxies, split between Seyfert AGN (black dotted), LINER AGN (green dash-dot), star
    forming (including composite, blue solid) and quiescent (red dashed).
    The fractional difference between observations and simulations is shown as a
    function of stellar mass. The comparisons with EAGLE (TNG100) are shown on the
    left (right) panels. From bottom to top, we show the lower (lGV), mid (mGV) and upper (uGV)
    green valley (see text for details). The error bars
    show the propagated Poisson uncertainty.}
  \label{fig:FracGalSDSS}  
\end{figure*}

\subsection{Fractional Variation in the Green Valley}
\label{sec:FracVar}

Although we already find limited differences between the EAGLE and TNG100
simulations in the BC, we find significant
difference in the RS, so we focus on the analysis of the GV region,
as it is a transition region where quenching
processes are expected to leave stronger imprints,
and where different feedback
prescriptions are expected to be more prominent.
Note that we start with $\sim 226$\, $872, 13$ \,475 and $22$\,$232$
SDSS, EAGLE and TNG100 galaxies, respectively with
$\log\, \mathrm{M_\star/M_\odot} \gtrsim 9.0$. After homogenisation
we are left with $88\,588$ (39.05$\%$ of total) EAGLE-SDSS and
$5\,822$ (43.21$\%$ of total) EAGLE galaxies
and $90\,709$ (39.98$\%$ of total) TNG100-SDSS and
$9\,906$ (44.56$\%$ of total)
TNG100 galaxies. 
The number of galaxies is further
reduced to $630$ (10.82$\%$ of homogenised sample) for EAGLE and
$486$ (4.91$\%$ of homogenised sample) for TNG100 when considering only
the GV. For the SDSS-EAGLE and SDSS-IllustrisTNG samples,
we find $6\,491$ (7.32$\%$) and $6\,448$ (7.11$\%$) galaxies, 
respectively in the GV. Even though there are similar fractions of EAGLE
and TNG100 galaxies after homogenisation, there is a greater drop
in TNG100 compared to EAGLE. Using these GV galaxies we
then compare the
fraction of different types of galaxies regarding nebular emission (as
defined in Sec.~\ref{sec:NebClassy}), in the lGV, mGV and uGV. The
classification separates galaxies into either SF, Q, LINER or Seyfert
AGN.  Fig.~\ref{fig:FracGalSDSS} compares the 
fraction of galaxies segregated by type as a 
function of stellar mass in EAGLE (${\it left}$) and TNG100
(${\it right}$), with respect to the observational constraints from SDSS.
The horizontal dashed line at zero represents the
ideal case where simulations and observations match perfectly.
The top, middle and bottom panels correspond to the uGV,
mGV and lGV, respectively. The lines are colour coded, as labelled.
In lGV and mGV, EAGLE is able to reproduce the general trend 
with respect to SF and Q galaxies at low stellar mass; as most data
straddle the zero point. However, for most massive galaxies
$>10^{10.8}$M$_\odot$,
EAGLE tends to overproduce (underproduce) quiescent (SF) galaxies. In uGV, 
the result is noisier but it is able to reproduce the fractions 
of observed galaxies. Note for low mass $<10^{10.5}$M$_\odot$,
the fraction suggests an overproduction of SF galaxies in 
simulations. In addition, there is a consistent underproduction 
of Seyfert AGN galaxies, however the samples 
include very little Seyfert AGN galaxies, so the statistics is not
so significant. The trend for LINER galaxies is noisier, so we cannot
deduce any mass trend of the mismatch with respect to the observations.

TNG100 also shows a similar level of
mismatch with respect to the
observational data, with an overproduction (underproduction) of Q (SF)
GV galaxies. The overproduction of quiescent galaxies 
seems to increase with respect to stellar mass, in both lGV and mGV.
Similar to EAGLE, there is a reasonable agreement between SDSS and 
TNG100 in the uGV. Note in lGV and mGV, for the most massive 
galaxies, $> 10^{10.5}$M$_\odot$, we find an underproduction 
of SF galaxies. Whereas in the uGV, 
TNG100 seems consistent with the observations. The fraction of Seyfert AGN
shows little difference. However this is once more owing to low
number statistics. The fractions for LINER is consistent 
with the observations as it straddles the zero point, similar
to EAGLE but noisier. 

There could be many reasons for the discrepancies found between the GV
galaxies in observations and simulations. We explore them in
Sec.~\ref{Sec:DnC}. Note the overabundance of Q galaxies, along with
the lack of SF galaxies in the simulation GV may indicate too rapid
quenching with respect to the observations. This is consistent with
the fact that both EAGLE and TNG100 feature a RS with a greater
4000\AA\ break strength than the SDSS observations.

\subsection{Specific Star Formation Rate}

\begin{figure*}
  \centering
  \includegraphics[width=82mm]{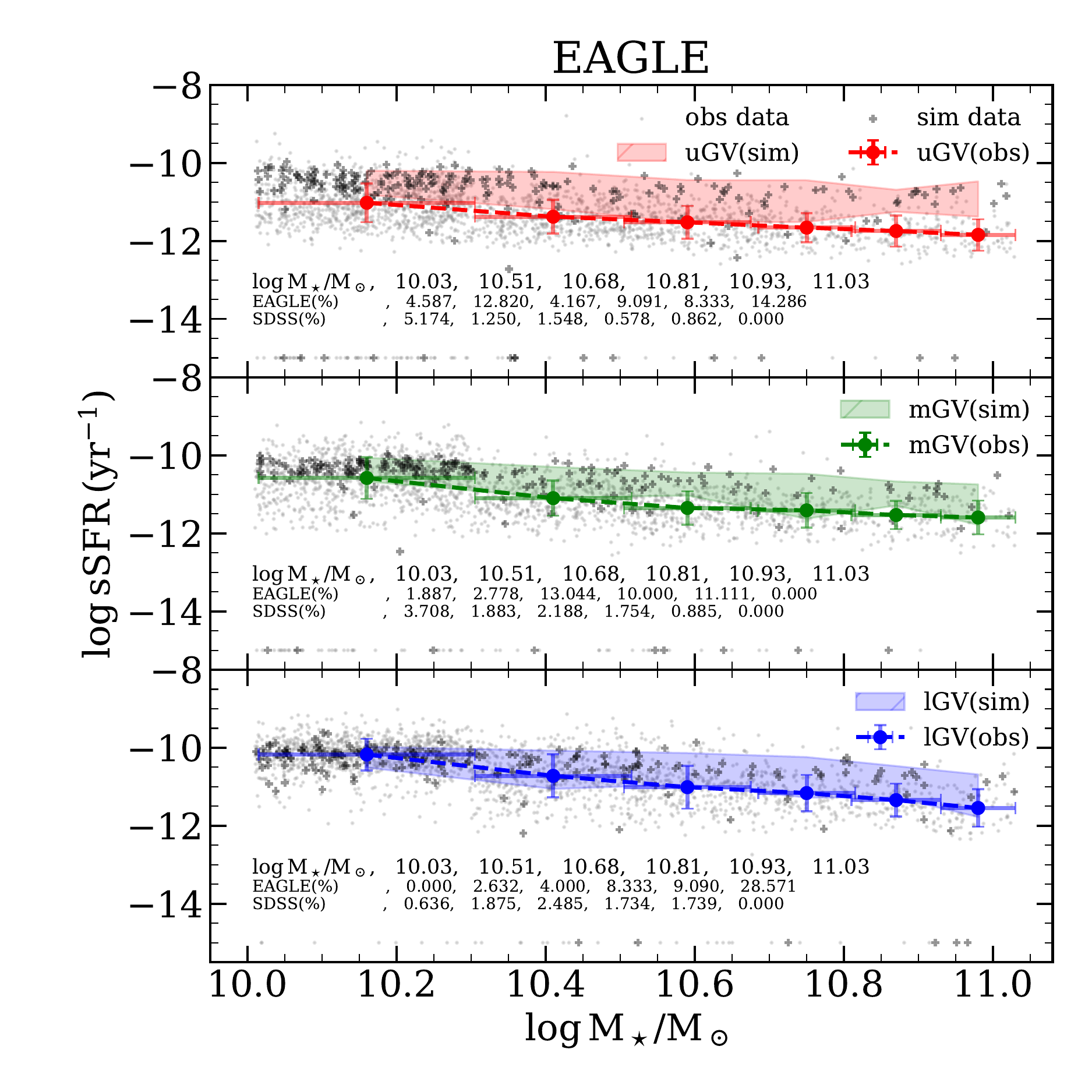}
  \includegraphics[width=82mm]{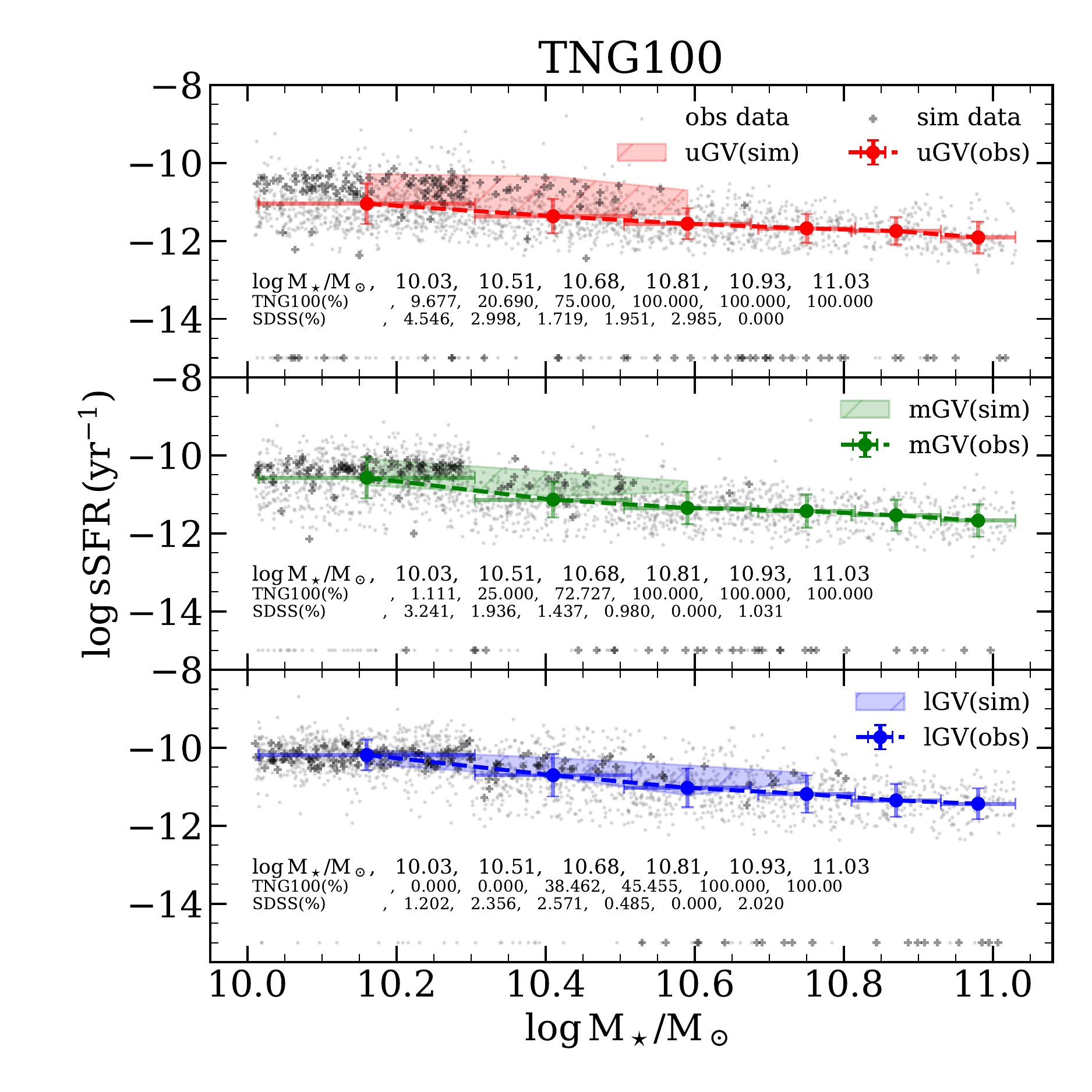}
  \caption{Distribution of GV galaxies on the sSFR vs stellar mass
  plane, The observations (simulations) are shown as gray (black)
  dots, with EAGLE (TNG100) simulations presented in the left (right)
  panel.  The blue, green and red dotted lines show the observational
  constraints of the lGV, mGV and uGV (bottom to top), and the 
  error bar is given at one standard deviation. The blue, green and red
  coloured sections show one standard deviation covered by the
  lGV, mGV and uGV from the simulations, respectively.
  Note the mean and standard
  deviation excludes all galaxies with zero SFR, i.e. following the
  convention log\,sSFR(yr$^{-1}$)=$- 15.5$, 
  to remove any systematics on the simulations 
  due to the numerical resolution limit. The text labels show the different
  stellar mass bins and the percentage of galaxies which have 
  their sSFR set to $-$15.5 for simulations and observations.}
  \label{fig:sSFR}  
\end{figure*}

In addition to the comparison based on the BPT classification of the
nebular emission lines, presented above, it is possible to further
test the models by comparing the behaviour of the specific star
formation (sSFR) rate, defined as the instantaneous star formation rate per unit
stellar mass.  sSFR is a powerful indicator of the ongoing stellar
mass growth.  Fig.~\ref{fig:sSFR} (left) shows sSFR against stellar
mass.  The blue, green and red dashed lines with error bars show the
mean observational sSFR in the lGV, mGV and uGV,
respectively, whereas the shaded regions delimit the results from the simulations.
Note for galaxies with SFR$=0$, we set their $\log\,$sSFR(yr$^{-1}$) at 
$-15.5$, and these galaxies are excluded when calculating the mean 
sSFR for each stellar mass bin. 
The black and gray data points correspond to individual galaxies in
the GV for simulations and SDSS, respectively. Galaxies with
$\log\,$sSFR$<-14$ (hereafter defined in yr$^{-1}$) represent fully quiescent
systems, i.e. with negligible star formation. 
The text labels on the figures state the
percentage of galaxies with SFR=0, thus have had their sSFR set to  
$\log\, \mathrm{sSFR} \equiv -15.5$, for their respective stellar
mass bins. Note these galaxies might have residual star formation, 
undetectable due to the resolution limit of simulations. However, 
the minimum SFR we find is $0.00071$ and $0.00032$ M$_\odot$\,yr$^{-1}$    
for EAGLE and TNG100, respectively. These estimates roughly translate to 
$\log\, \mathrm{sSFR} = -14.2$ for the most massive galaxies. Therefore,
even without the limit in resolution regarding the star formation histories, 
these galaxies would
have a lower SFR than the SDSS sample. Note, for observational
galaxies with undetectable star formation, their sSFR have artificially
been put around $\log\, \mathrm{sSFR} = -12.0$ following a gaussian 
distribution \citep{Jarle:04,Don:19}. We argue, since we are looking in 
GV and in stellar mass range, 
$\log\, \mathrm{M_\star/M_\odot} = 10^{10.03-11.03}$, 
that the fraction of galaxies
with undetectable star formation should be low. 
This is further backed up by the low fraction 
of Q galaxies present in the GV -- see Sec. \ref{Sec:DnC_SFH}.
EAGLE produces a trend of sSFR
with stellar mass that is close to the observations. The
SDSS data feature a wider separation between lGV and uGV, whereas
the simulations show significant overlap. Moreover, the
simulations show overall higher values of the sSFR with respect to the
observations. This mismatch may be due to a systematic offset,
as the sSFR of the simulations is retrieved from the output of the star
formation activity, whereas the observational constraints are
determined indirectly from standard relations involving emission
lines \citep[see, e.g.,][]{RobK:98}. However, 
whilst simulation galaxies have 
higher SFR in the star-forming subsample, they produce a higher fraction 
of galaxies with no star formation, thus
this trend is consistent with our analysis based on the fraction of 
Q and SF galaxies, shown in the previous subsection.

The right hand panel of Fig.~\ref{fig:sSFR} corresponds to the TNG100
simulation. The overall trend with respect to stellar mass is
consistent, but the simulations produce values that are systematically
higher, and the mismatch is similar with EAGLE galaxies.
While the SFR is systematically higher, we do
find a higher fraction of galaxies, in TNG100, with no ongoing
star formation. Furthermore, at the
massive end, $\log\,$M$_\star$/M$_{\odot}\gtrsim 10.6$, most
TNG100 GV galaxies lack star formation whereas the observations
feature a significant number of SF galaxies, once more supporting the argument
towards an excess of Q galaxies in the simulations at the massive
end.

Even though we are comparing galaxies 
with similar D$_n$(4000) breaks, the simulated galaxies happen to 
be lower on the evolutionary sequence in comparison to observations
(Fig.~\ref{fig:GV_D4k_Sim}). Therefore, the higher sSFR found 
for the star forming galaxies is somewhat expected.

\subsection{Average Ages}
\label{sec:Ave_time}

The next step in the comparison of GV galaxies involves the comparison of the
trend in average age, between observations and simulations.
For the observational (SDSS) data, we adopt the same 
procedure as in A19, and A20, stacking the spectra, following 
\citet{Ferreras:13}. The stacks are presented to \verb|STARLIGHT| \citep{SLight},
to extract star formation histories. This code performs 
full spectral fitting with an MCMC-based algorithm that finds the
best-fit weights of a set of simple stellar populations (SSP).
From these weights the average age is calculated as follows:

\begin{equation}
      \langle\log\, t\rangle \equiv \sum^{N_\star}_{j=1} x_j \log\, t_j,
      \label{eq:SL_tave}
\end{equation}
where $x_j$ is the normalised luminosity weight and $t_j$ is the
stellar age corresponding to the $j$-th SSP.
For spectral fitting we use 138 SSPs from the \citet{BC03} models, where 
the age
varies from 0.001 to 13\,Gyr and total metallicity varies from $10^{-4}$ 
to 0.05.
The ages in the simulated galaxies are determined directly from the
stellar growth of the galaxies, i.e. by taking into account the
distribution of stellar particles within a R=3\,kpc aperture size; to 
better
match the SDSS classic spectra, taken through optical fibres that map 
a 3\,arcsec diameter. Analogously to the spectral
stacking performed in the SDSS data, the star formation histories for
each galaxy within the same bin (in stellar mass and location on the
GV) are stacked to create a joint distribution of stellar ages from
which the average is determined.  The uncertainties in both cases are
obtained from a bootstrap, where each realization randomly stacks 60\%
of the galaxies in each stellar mass bin. 
The uncertainties of average stellar ages 
vary between $\sim 0.1 - 0.7$\,Gyr for observational constraints,
from \texttt{STARLIGHT}, 
and  $\sim 0.3 - 0.5$\,Gyr for simulated galaxies. 
Note that the stellar ages
can be weighed either by luminosity or by mass. The former are better
constrained by the spectra, whereas the latter are more physically
motivated. Similarly to the dilemma between velocity dispersion and
stellar mass -- where the former is better constrained by the
observations, and the latter is the preferred choice for simulations,
in this case luminosity-weighted ages are more accurately constrained
by the observational spectra, whereas mass-weighted ages suffer less
systematic in the simulations.

\begin{figure*}
  \centering
  \includegraphics[width=80mm, height=80mm]{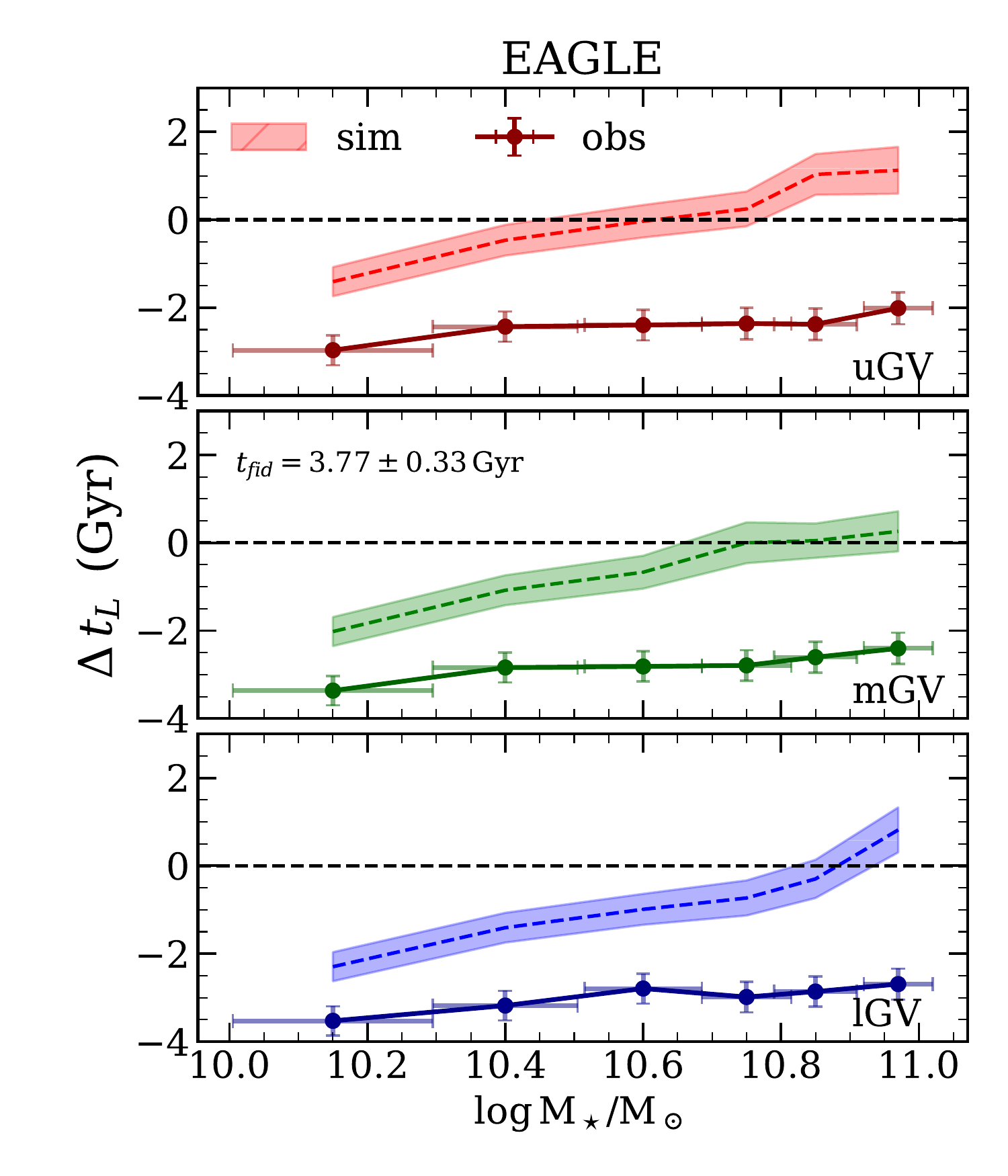}
  \includegraphics[width=80mm, height=80mm]{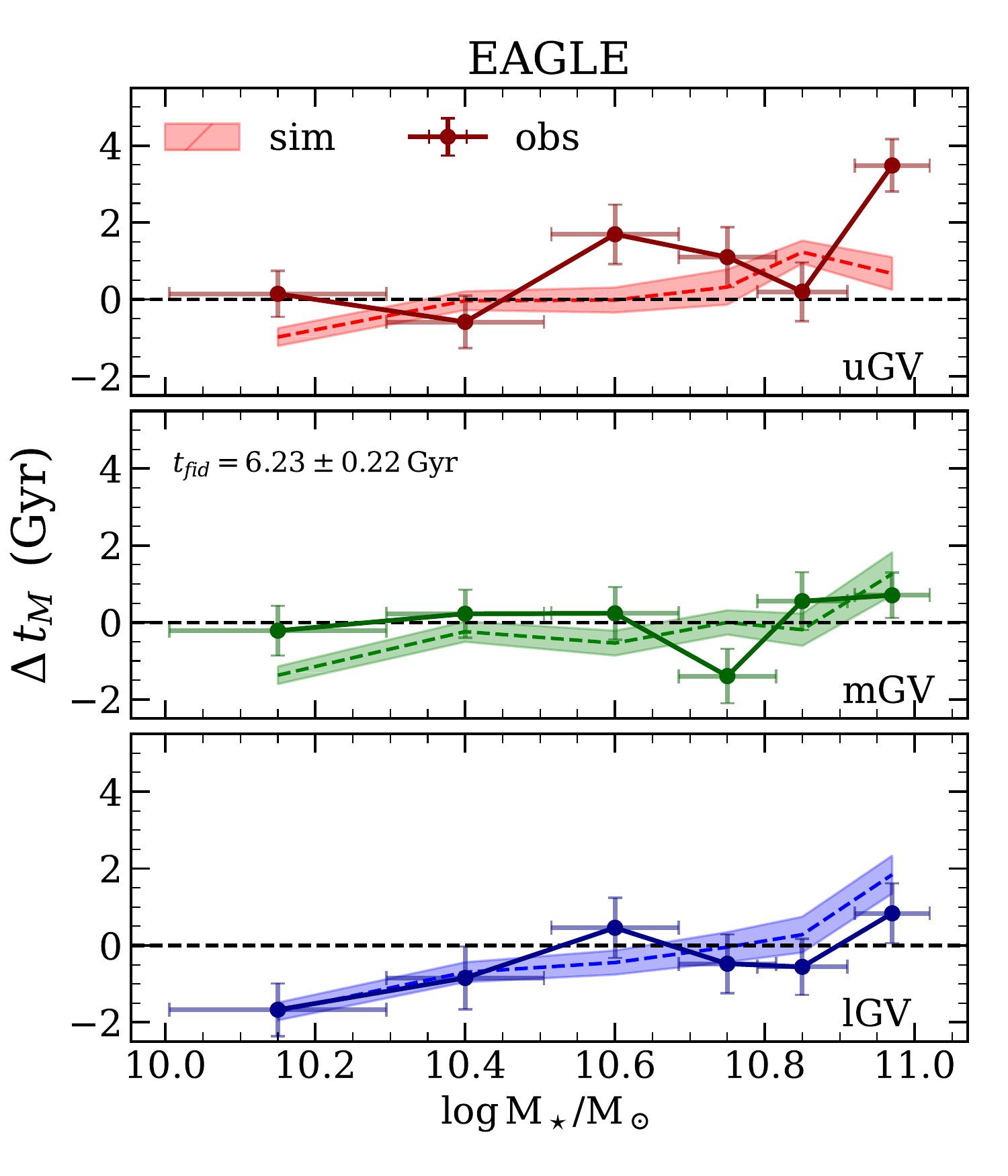}
  \includegraphics[width=80mm, height=80mm]{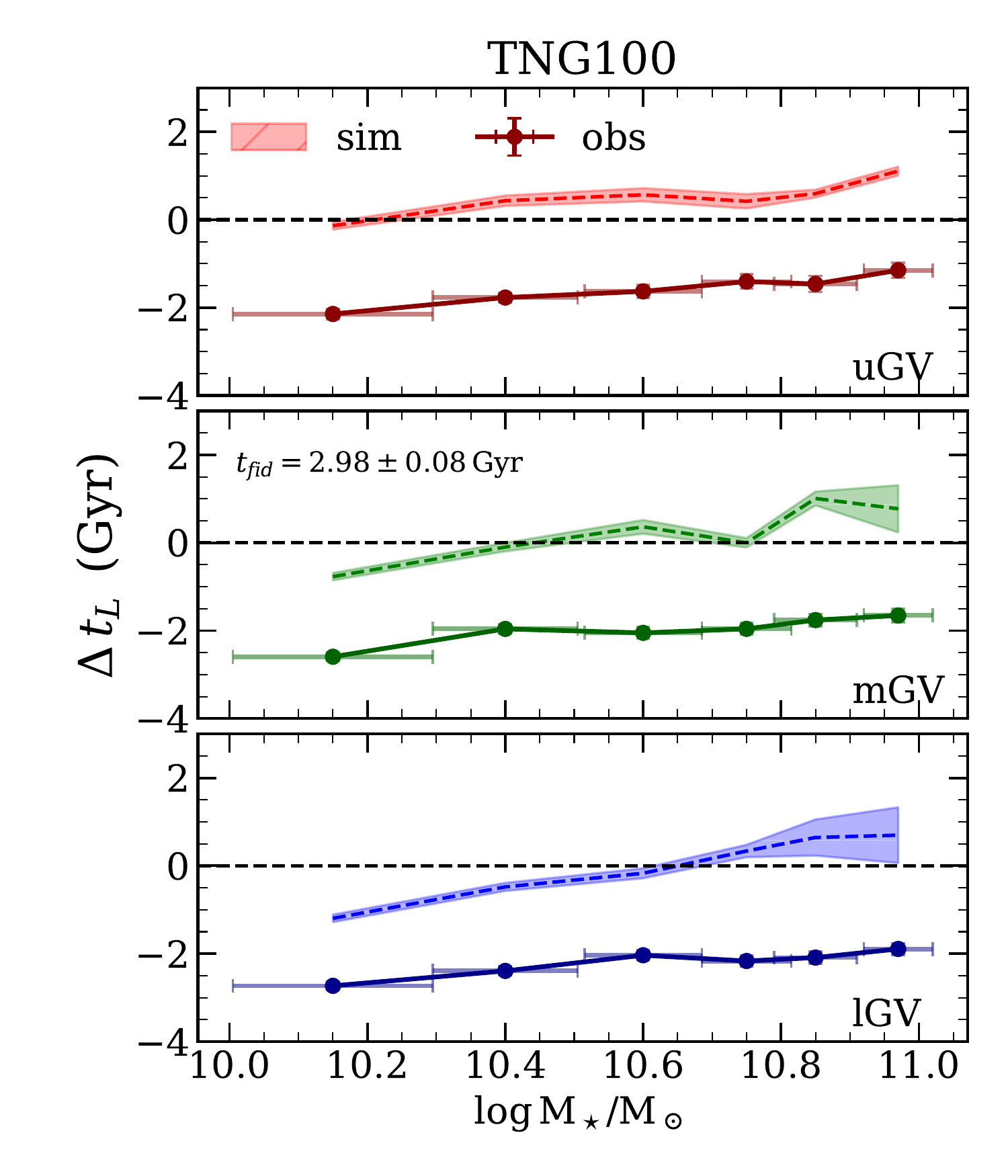}
  \includegraphics[width=80mm, height=80mm]{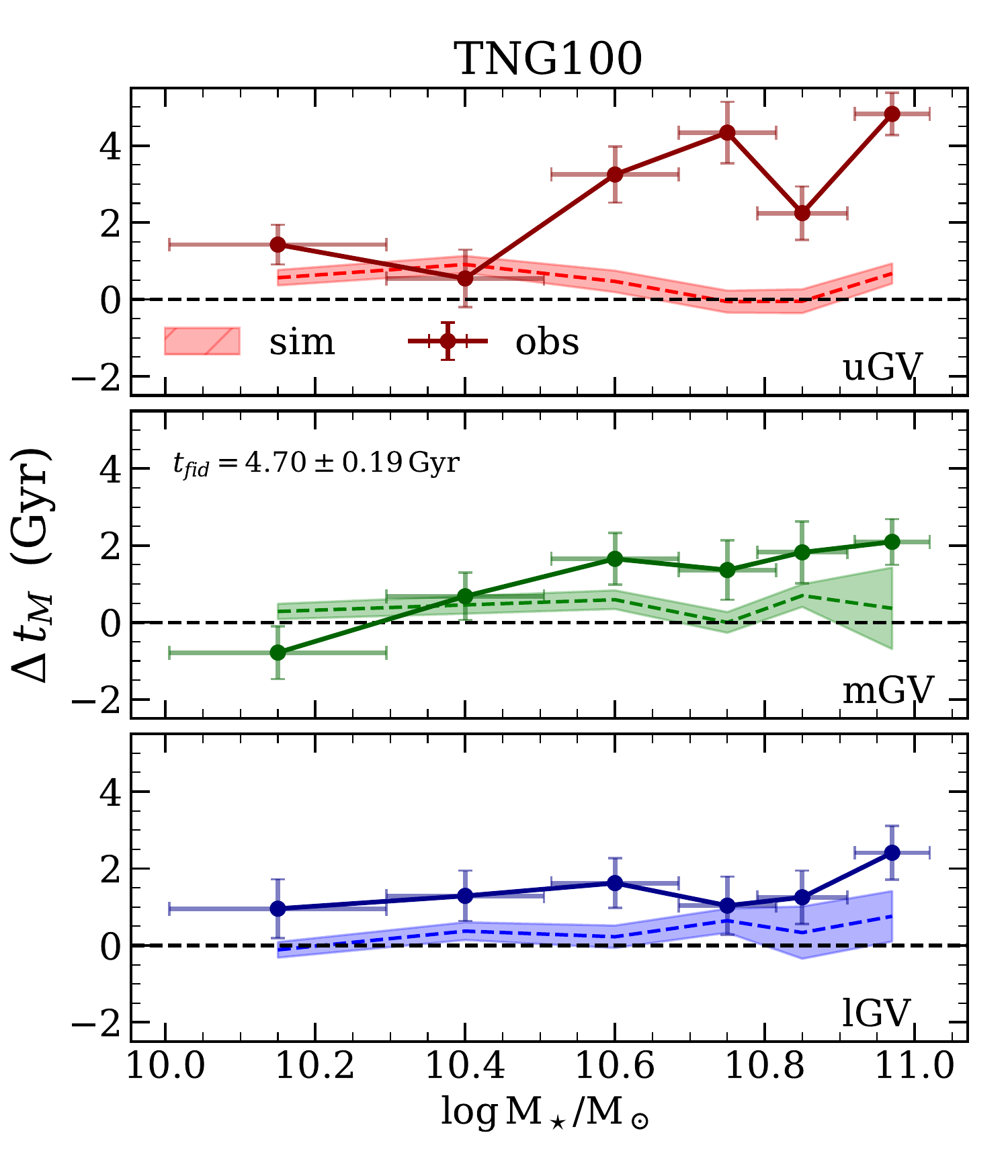}
  \caption{Relative luminosity- ({\it left}, $\Delta t_L$) and mass-
  ({\it right}, $\Delta t_M$) weighted
  average stellar ages, shown with respect to stellar mass. Comparison
  with EAGLE (TNG100) are shown in the top (bottom) figures, each one separated
  into panels that correspond (from bottom to top) to the lGV, mGV, uGV.
  The relative ages are measured with respect to a fiducial one ($t_{\rm fid}$) 
  corresponding to mGV galaxies in the 
  10.68$\lesssim\log\,$M$_\star$/M$_\odot$$\lesssim$10.81 mass bin of the
  simulation data, quoted in each figure (see text for details).}
  \label{fig:Ave_Age_Params}  
\end{figure*}

Taking into account that relative age variations are more robustly
constrained than absolute estimates, we decide to quote our results
as a relative age difference, given by:
\begin{equation}
        \Delta \psi_k(t) = t_i^k - t_{\rm fid},
        \label{eq:Rel_diff}
\end{equation}
where $\psi$ represents either the average age or quenching timescale
(see below). The index $k$=$\{L,M\}$ denotes whether the parameter is
luminosity- or mass-weighted, while $i$ denotes the chosen stellar
mass bin. Note for each of the four parameters explored -- namely
luminosity-weighted average age and quenching timescale, and
mass-weighted average age and quenching timescale -- we select a
single fiducial value throughout, defined as the estimate from
simulations in the mGV, at mass bin
$10^{10.68}<$M$_\star$/M$_\odot<10^{10.81}$. We then subtract this
fiducial value, both in simulation and observation parameters, across
all stellar mass bins and GV regions. Therefore, by construction,
$\Delta\psi_k$ for the fiducial bin will be zero for simulations,
whereas the value of $\Delta\psi_k$ in this fiducial bin for the
observed data will account for systematic offsets between observations
and simulations.

Fig.~\ref{fig:Ave_Age_Params} shows the relative average stellar ages
in the observations (solid lines) and the simulations (filled dashed
lines), with EAGLE (TNG100) shown in the top (bottom) panels. We also
show separately the luminosity- (left) and mass-weighted (right)
values.  In each case, the top, mid and bottom panels show the result
of uGV, mGV and lGV, respectively. The values quoted in the
mid panel show the actual
estimate of the stellar age corresponding to the fiducial mass
bin. The luminosity-weighted ages show an increasing trend with
stellar mass in all cases, consistent with the well-established
mass-age relation \citep{Kauff:03, Gallazzi:05}.  In all three regions
of the GV, EAGLE produces galaxies that are both older, in luminosity
weights, than the
observed constraints (by about 2.8\,Gyr at the fiducial bin), and with
a steeper mass-age slope (see Tab.~\ref{tab:slopes_t}).
The luminosity-weighted estimates show a substantial systematic 
between
the spectral fitting results (i.e. the data points) and the
constraints from the star formation histories of the cosmological
simulations (i.e. the shaded regions), in all GV regions and in both
EAGLE and TNG100. There is a slightly better agreement with 
observations in the slope of the luminosity-weighted estimates of
TNG100, (see Tab.~\ref{tab:slopes_t}), and the fiducial age is also
slightly closer to the observational constraints ($\sim$2.0\,Gyr).

\begin{table*}
  \centering
  \caption{Slopes of the relation between average age and stellar mass.
  The slope ($\alpha$) is obtained from a
  linear fit to the function $\Delta t_{\rm L,M}=\alpha\log$\,M$_\star$/M$_\odot + \beta$
  (see Fig.~\ref{fig:Ave_Age_Params}).
  The uncertainty is quoted at the 1\,$\sigma$ level.}
  \label{tab:slopes_t}
    \begin{tabular}{ccccc} 
    \hline
     & \multicolumn{2}{c}{Lum-weighted} & \multicolumn{2}{c}{Mass-weighted}\\
     & EAGLE & SDSS & EAGLE & SDSS\\
    \hline
    lGV & $+3.31\pm 0.49$ & $+0.92\pm 0.20$ & $+3.62\pm 0.72$ & $+2.31\pm 0.95$\\
    mGV & $+2.82\pm 0.22$ & $+0.99\pm 0.18$ & $+2.40\pm 0.74$ & $+0.52\pm 1.23$\\
    uGV & $+3.09\pm 0.27$ & $+0.91\pm 0.22$ & $+2.23\pm 0.52$ & $+3.11\pm 1.81$\\
    \hline
     & \multicolumn{2}{c}{Lum-weighted} & \multicolumn{2}{c}{Mass-weighted}\\
     & TNG100 & SDSS & TNG100 & SDSS\\
     \hline
    lGV & $+2.40\pm 0.16$ & $+0.92\pm 0.18$ & $+0.85\pm 0.28$ & $+1.08\pm 0.69$\\
    mGV & $+1.90\pm 0.47$ & $+0.98\pm 0.22$ & $+0.09\pm 0.40$ & $+3.28\pm 0.59$\\
    uGV & $+1.16\pm 0.31$ & $+1.10\pm 0.11$ & $-0.56\pm 0.58$ & $+4.14\pm 1.80$\\
    \hline
    \end{tabular}
\end{table*}

In both cases, the mass-weighted average ages show better agreement
with observations, both regarding the systematic offset and the
slope, although with  larger error bars.
However, note the discrepancy at the massive end of the TNG100
uGV, where simulated galaxies have substantially older ages.
Note previous work hinted the upper part of the GV could be the
least homogeneous, as one could have a more complex mixture of
galaxies evolving from the BC to the RS via quenching, as well
as galaxies moving ``backwards'' due to mild, but frequent
episodes of rejuvenation \citep[][A20]{Rejuv2010,Nel:18}. 

\begin{table*}
  \centering
  \caption{Equivalent of Table~\ref{tab:slopes_t} for the quenching timescale,
  $\tau_Q$. The slope ($\alpha$) is obtained from a
  linear fit to the function $\Delta \tau_{\rm Q;L,M}=\alpha\log$\,M$_\star$/M$_\odot + \beta$
  (see Fig.~\ref{fig:tQ_Age_Params}).
  The uncertainty is quoted at the 1\,$\sigma$ level.}
  \label{tab:slopes_tauQ}
    \begin{tabular}{ccccc} 
    \hline
     & \multicolumn{2}{c}{Lum-weighted} & \multicolumn{2}{c}{Mass-weighted}\\
     & EAGLE & SDSS & EAGLE & SDSS\\
    \hline
    lGV & $+1.32\pm 0.96$ & $-0.74\pm 1.16$ & $-1.25\pm 0.70$ & $-1.92\pm 1.17$\\
    mGV & $+0.71\pm 0.98$ & $-0.03\pm 0.64$ & $+0.30\pm 0.48$ & $-1.83\pm 1.31$\\
    uGV & $-1.08\pm 1.36$ & $-0.85\pm 1.33$ & $-1.38\pm 0.68$ & $-8.04\pm 2.12$\\
    \hline
     & \multicolumn{2}{c}{Lum-weighted} & \multicolumn{2}{c}{Mass-weighted}\\
     & TNG100 & SDSS & TNG100 & SDSS\\
     \hline
    lGV & $+1.61\pm 0.72$ & $-0.68\pm 1.12$ & $-0.40\pm 0.89$ & $-1.31\pm 1.66$\\
    mGV & $-0.13\pm 0.78$ & $-0.20\pm 1.35$ & $+0.07\pm 0.72$ & $-0.20\pm 3.17$\\
    uGV & $-1.33\pm 0.92$ & $-0.27\pm 2.06$ & $-0.69\pm 0.77$ & $+1.85\pm 2.09$\\
    \hline
    \end{tabular}
\end{table*}

\begin{figure*}
  \centering
  \includegraphics[width=80mm, height=80mm]{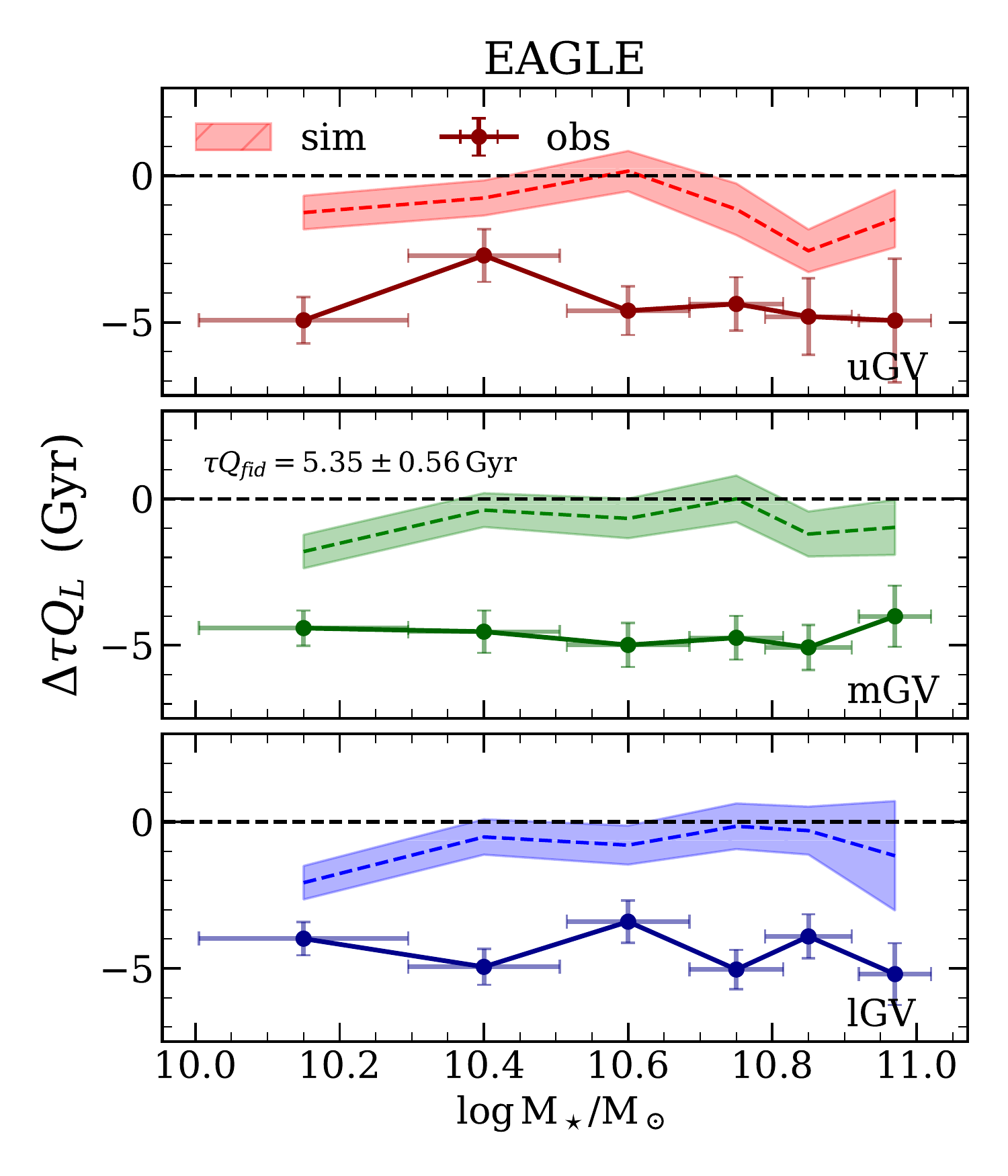}
  \includegraphics[width=80mm, height=80mm]{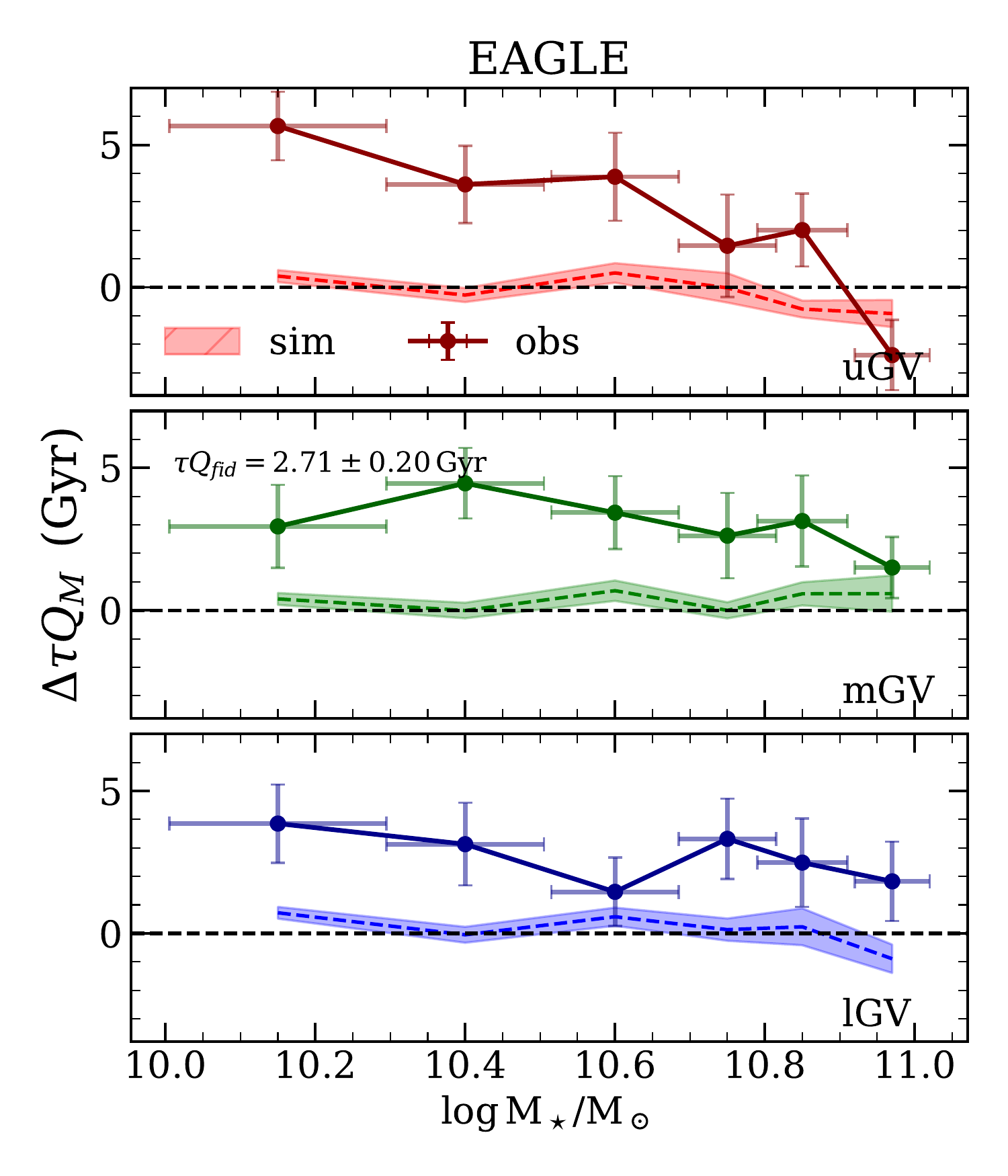}
  \includegraphics[width=80mm, height=80mm]{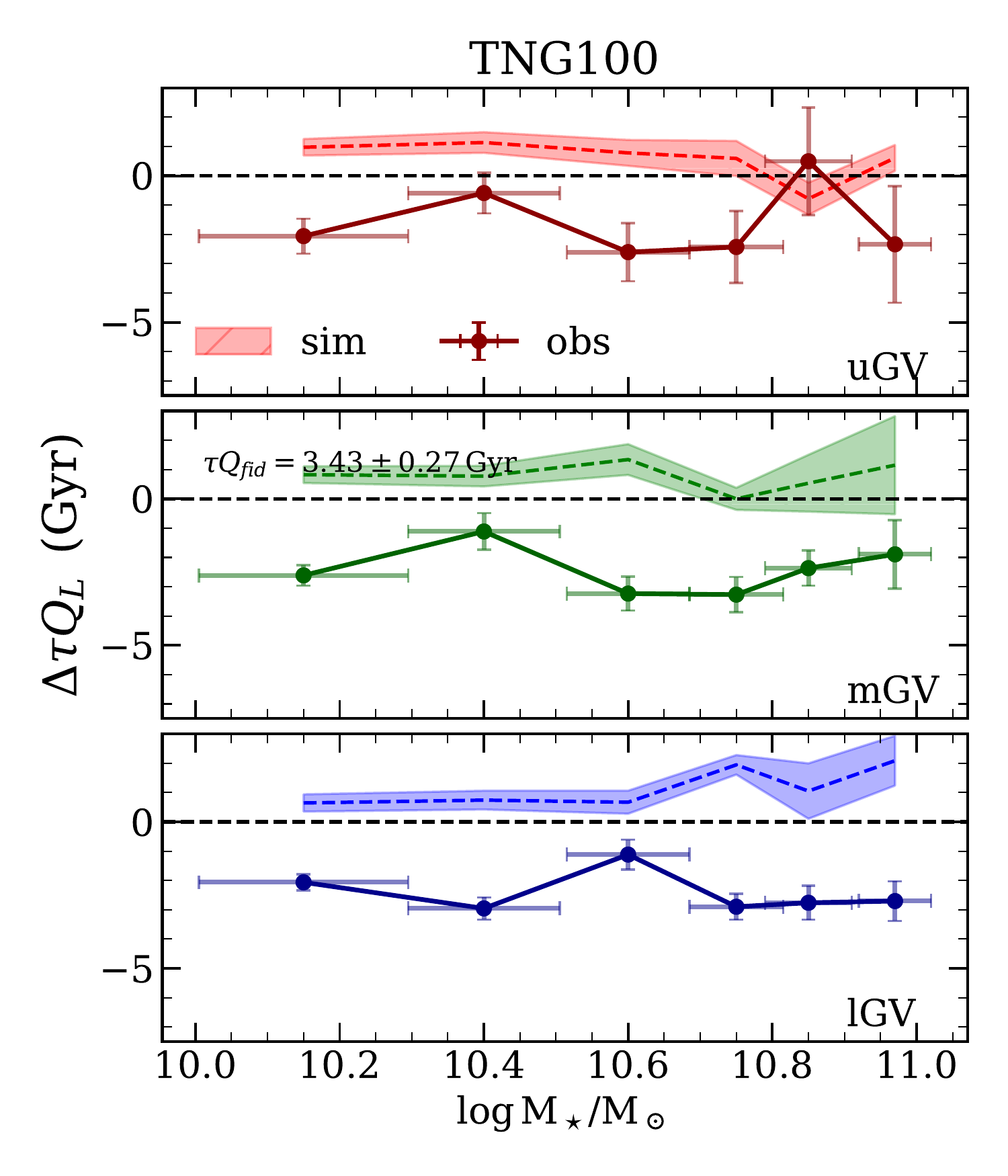}
  \includegraphics[width=80mm, height=80mm]{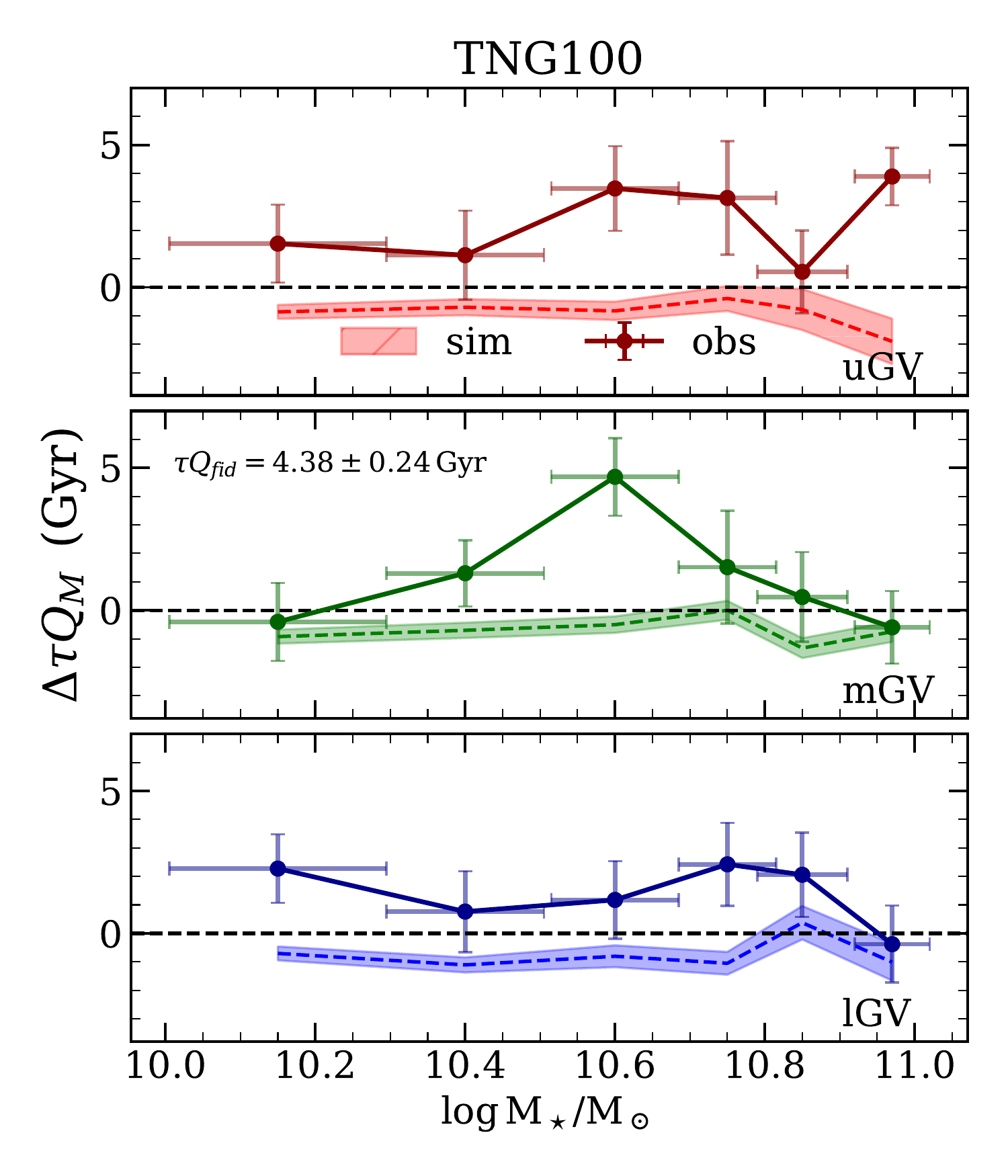} 
  \caption{Equivalent of Fig.~\ref{fig:Ave_Age_Params} for the
  relative quenching timescale, $\Delta\tau_Q$, quoted with
  respect to a fiducial value ($\tau_{\rm Q,fid}$), also taken 
  from mGV galaxies in the 
  10.68$\lesssim\log\,$M$_\star$/M$_\odot$$\lesssim$10.81 mass bin of the
  simulation data, and quoted in each figure.}
  \label{fig:tQ_Age_Params}  
\end{figure*}

\subsection{Quenching timescales}
\label{sec:Quench_time}

In A20, we presented a parameter that -- under a number of simplifying
assumptions -- can serve as a proxy of the quenching timescale. It is
defined as the time interval between two percentile levels of the
stellar age distribution, namely:
\begin{equation}
     \tau_Q \equiv t_{70} - t_{30},
     \label{eq:SL_tQ}
\end{equation}
where $t_x$ represents the cosmic time when the cumulative stellar mass
function reaches a percentile level of $x$. Therefore $\tau_Q$ is the
time that the system takes to go from a stellar mass content of 30\% to
70\% of the final amount. 
Similarly to average age estimates we bootstrap 60\%
of the galaxies to obtain uncertainties in the range
$\sim 0.07$-$2.3$\, Gyr for the observational data, derived from
\texttt{STARLIGHT}, and $\sim 0.1$-$0.8$\, Gyr
for simulated data, directly from the star formation histories.
In a simple scenario where the galaxy builds up
the stellar content in a monotonic way, this parameter scales with the
rate at which stellar mass grows. Ideally, one would consider higher 
levels of the percentile (i.e. $x$), in order to represent more 
accurately the final stages before quenching ensues. 
However, our choice is motivated by the unavoidable
uncertainties of the analysis, notwithstanding the systematics related to
the derivation of the star formation history from full spectral fitting,
following the {\sc Starlight} code \citep{SLight}. 

Fig.~\ref{fig:tQ_Age_Params} shows the relative trends in both
luminosity- (left) and mass-weighted (right) quenching timescale. The
symbols and panels are analogous to those presented in
Fig.~\ref{fig:Ave_Age_Params}, comparing the results for EAGLE and TNG100
in the top and bottom panels, respectively. The slopes of these 
relations,
assuming a simple linear trend, are quantified in 
Table~\ref{tab:slopes_tauQ}.
We emphasize that
$\tau_Q$ is less robust than the derivation of an average case, as
presented above. However, we note that the constraints are imposed on
stacked spectra of galaxies with very similar properties (same stellar
mass and in the same GV region), and feature a high signal to noise
ratio. Therefore, the {\sl relative} variations in $\tau_Q$ among the
stacked data are more reliable than individual measurements of the
same parameter. We also note that, similarly to the average age shown
above, we present the data as relative to a fiducial bin, chosen to be 
$10^{10.68}<$M$_\star$/M$_\odot<10^{10.81}$ interval in mGV simulated 
galaxies.
Two fundamental pieces of information should be looked for in the
figures: the general trend with stellar mass, and the overall offset
between SDSS data and simulations.
In the luminosity-weighted case (left panels), both simulations 
reproduce the
general trend of the observations, with an overall flat slope.
There is a substantial offset between both, with SDSS data featuring
shorter values of $\tau_Q$, analogous to the systematic in average age.
Interestingly, this offset is reversed in the mass-weighted case, where
the comparison suggests that the simulations quench star formation more
rapidly. Moreover, this behaviour also appears in the TNG100 comparison.

Note the sign reversal of $\Delta\tau_Q$ between the
luminosity-weighted (left) and the mass-weighted cases (right), that
appears in both EAGLE and TNG100. This sign change implies the
observations produce shorter $\tau_Q$ than the simulations when
luminosity-weighted, but longer values of $\tau_Q$ when weighed by
mass. This result could be caused by the fact that the 
D$_n$(4000) index is sensitive to the presence of 
younger stellar populations from a recent episode of star formation
\citep{D4000ref}. However, this could also be explained by the way
{\sc STARLIGHT} constrains the weights of individual SSPs when
performing spectral fitting. Luminosity weighting puts high weights
in a short and recent period of time. Hence the resulting $\tau Q$ is 
short.
In contrast, the mass-weighted SFH distributes evenly the weights of
the different stellar populations, producing a more extended distribution
that results in a longer $\tau_Q$.
Simulations distribute the weight of the individual star particles
evenly for all ages, hence is not so sensitive to the difference between
luminosity and mass weighing.

\section{Discussion}
\label{Sec:DnC}
This paper focuses on a recent classification of GV galaxies,
following a probability-based methodology applied to the distribution
of 4000\AA\ break strength in a large sample of SDSS (classic) spectra
with relatively high signal-to-noise ratio. The definition of the GV
along with three sub-regions (lower-, mid-, and upper-GV, denoted lGV,
mGV, uGV, respectively) was presented and analysed in A19 and A20.  We
contrast here the observational properties presented in those papers,
with two state-of-the-art cosmological simulations, EAGLE and
Illustris TNG. Since the GV can be interpreted as a transition region
where galaxies evolve from actively star-forming to quiescent systems,
it should be considered a fundamental sample where the physical
mechanisms of feedback, described in cosmological simulations by an
overly simplified set of equations, loosely termed subgrid physics, can be put
to the test. In a way, constraints based on GV galaxies provide ``the
next order'' in our perturbative approach towards galaxy formation,
the lowest order being the standard constraints on the galaxy
luminosity/mass function, the Tully-Fisher relation and the
mass-metallicity relation \citep{Schaye:15, Pillepich:18}.

\subsection{Potential Caveats}

\subsubsection{Population synthesis models}
In order to follow the same definition of the GV in the simulations
as in A19 and A20,
we had to create synthetic spectra by combining the star formation and
chemical enrichment history of the simulations with the E-MILES 
population synthesis models of \citet{eMILES}. Once the mock spectra
were created, we measured the 4000\AA\ break strength following the
same D$_n$(4000) index \citep{Balogh:99}. A possible systematic lies
in the choice of models, including stellar evolution/isochrone prescriptions
and stellar libraries. One would consider contrasting the results
with respect to independent population synthesis models, such as
BC03 \citep{BC03} or FSPS \citep{FSPS:2010}, beyond the scope of this
paper. We note, though, that our analysis is mostly focused on {\sl relative}
variations of the stellar ages, which present significantly smaller systematics
than {\sl absolute} age estimates. Moreover, our population analysis work
is consistent with the comparisons based on nebular emission alone, a
simpler and robust characterization of the sample.

\subsubsection{Green Valley definition}
\label{Sec:DnC_GV}

An additional caveat regarding the mismatch  between simulations 
and observations may be due to the actual definition of the GV
distribution function. Previous
studies of the EAGLE and TNG100 samples selected GV galaxies based
on colour, or chose the GV on the SFR vs stellar mass plane
\citep{Tray:15, Tray:17, Wright:18, Nel:18, Correa:19}. 
Our selection, based on 4000\AA\ break strength is motivated by a
fully empirical approach, and requires the ``projection'' of the
output from the simulations on the observational plane by creating
synthetic spectra (Sec.~\ref{sec:SynthSpec}) whose D$_n$(4000) is
measured.  This step may introduce a systematic that can affect the
comparison.  An alternative approach would require defining the GV
in the simulations independently, similarly to 
\cite{Tray:15}, \cite{Wright:18} for EAGLE and \cite{Nel:18}
for TNG100, thus having independent definitions 
of the GV for EAGLE, TNG100 and SDSS. Using this methodology, 
we would have different GV locations for EAGLE, TNG100 and SDSS on the
D$_n$(4000) vs stellar mass plane. This would produce independent lGV,
mGV and uGV samples in each data set. However this method is
counterproductive to the aim of this paper, as we are aiming at
comparing the empirical definition of the GV, based on SDSS data, with
model predictions. Thus by projecting simulations onto the
observational plane and comparing the results enables the improvement
of the subgrid physics such that the models can reproduce the BC, GV
and RS morphology as found in the observations.  Moreover, by focusing
on the GV in this manner, we isolate the cause of the discrepancy
i.e. the overabundance of Q GV population and lack of SF GV galaxies
hinting towards over quenching or too rapid quenching in the
simulations.

\subsection{Contrasting observations with simulations}
\label{Sec:DnC_Study}

\subsubsection{Overall distribution}
Hydrodynamical simulations have been shown to successfully reproduce
the fundamental scaling relations of galaxies. Furthermore, there have
been multiple studies on the bimodality that find a qualitative
agreement between observation and simulations: in EAGLE, multiple studies
have tested
using the colour (g--r) at $z=0.1$, to find an overall agreement with
the distribution with observation \citep{Tray:15}. Implementing 
a more sophisticated treatment of dust attenuation --  where
younger stellar populations are dustier than older systems -- 
a better agreement with data from the GAMA survey was
obtained \citep{Tray:17}. Note these authors report a red
sequence that appears slightly flatter than observed, in agreement
with our Fig.~\ref{fig:GV_D4k_Sim}. The flatter red sequence gradient
was attributed to a flatter mass-metallicity relation \citep{Schaye:15}.
This discrepancy is also evident in the definition based on the 
D$_n$(4000) index, as it is substantially more sensitive to metallicity
than colour.

Concerning TNG100-based comparisons, \citet{Nel:18} correct for dust
attenuation, finding a good quantitative agreement of the bimodality
with the observations on the colour vs stellar mass plane.  We also
find good qualitative agreement on the colour vs stellar mass plane,
by applying a simple dust correction with the \citet{Calz:94} law.
However, our analysis based on the D$_n$(4000) index, which is
significantly less sensitive to dust attenuation, shows a greater
discrepancy (Fig.~\ref{fig:GV_D4k_Sim}), and this behaviour is similar
in the EAGLE simulations. This result would suggest that the
prescriptions chosen by the simulations, especially manifesting on the
mass-metallicity relation, are able to reproduce and explain the
observed colours, whereas the 4000\AA\ break selection of GV suggests
these prescriptions are not good enough to explain in detail the
properties of the underlying stellar populations. Note that both
broadband colours and D$_n$(4000) suffer from the age-metallicity
degeneracy \citep[see, e.g.][]{Worthey:94}, whereas the latter has a
negligible dependence on dust. Therefore, the subgrid physics needed
to constrain either may introduce independent biases. We argue here
that the 4000\,\AA\ break strength is a more fundamental observable,
as it removes the highly complex layer of dust production,
destruction, geometry, radiative transfer, etc, needed to produce
reliable estimates of broadband photometry.

\subsubsection{Overquenching}

We discuss here the potential explanation of the mismatch of the
bimodality shown with 4000\AA\ break strength as a result of
overquenching. Note most of our analysis is carried out on the GV.

Although qualitatively there is good agreement between the two
simulations, in EAGLE we find the RS features a D$_n$(4000)
index $\sim$0.1\,dex higher than the SDSS constraints at stellar
mass $\lesssim 10^{10.5}$M/M$_\odot$. TNG100 also produces a RS
with an even  greater D$_n$(4000) strength, $\sim$0.2\,dex higher than SDSS.
Both simulations produce the BC in agreement with the observations,
therefore indicating a mismatch to feedback/quenching mechanisms that
eventually produce the RS. This hypothesis is backed by the
fraction of galaxies classified with respect to nebular emission
(Fig.~\ref{fig:FracGalSDSS}) that shows higher fraction
of Q galaxies in the GV, especially at the massive end. Note
that the definition of quiescence versus star formation is less prone
to biases than the definition of AGN activity
(Tab.~\ref{tab:AGNdef}). Therefore, the excess of Q
GV galaxies in simulations appears to be quite robust.  Luminosity-weighted
average ages (Fig.~\ref{fig:Ave_Age_Params}), consistently show older galaxies
in the simulated systems in all GV regions,
lGV, mGV and uGV. Between the
two sets, TNG100 features the larger discrepancy towards a larger
excess of Q galaxies and older average ages. 
Fig.~\ref{fig:sSFR} shows the simulated GV
galaxies with slightly higher sSFR than the observations, but within 
1$\sigma$ in all lGV, mGV and uGV. The greatest difference between
simulations and observations is due to the former lacking any 
ongoing star formation, where EAGLE and TNG100 have $28 \%$ and 
$100\%$ GV galaxies with $\log\, \mathrm{sSFR} < -15.5$ at the highest stellar 
mass bins $\geq 10^{10.81} \mathrm{M_\odot}$. While EAGLE produces a 
consistent level of scatter, $0.5-0.8\,$dex, matching the observations, 
a lower scatter, $0.2-0.5\,$dex, is seen for TNG100 at
high stellar mass, $\geq 10^{10.68} \mathrm{M_\odot}$. 

Previous studies of GV galaxies in both EAGLE and TNG100 suggest 
quenching timescales 1.5$\lesssim \tau_{\rm GV}\lesssim$7.0\,Gyr
\citep{Tray:16,Wright:18,Nel:18,Correa:19}, a range that mostly
agrees with the constraints derived from SDSS and GAMA data
\citep[][A19, A20]{Schawinski:14, Phill:19}, 
with a heavy dependence on morphology. Our study of GV and quenching 
timescale also gives values within this limit in both observations 
and simulations, as shown in Fig.~\ref{fig:tQ_Age_Params}. However 
note our sample is less homogeneous than the previous studies, 
as we stack the SFHs regardless of galaxy type or morphology, thus
we are unable to find a clear trend with respect to stellar
mass. The mass-weighted quenching timescale supports overquenching/
rapid quenching as the primary reason for 
the discrepancy between observations and simulations: 
we obtain more extended SFHs for
SDSS than both EAGLE and TNG100, $\tau_Q({\rm obs}) > \tau_Q({\rm sim})$. 
Note both EAGLE and TNG100 reproduce the expected decreasing quenching timescale
with stellar mass,  as found 
in the literature \citep{Kauff:03, AnnaGallD4k}. However
for the luminosity-weighted quenching timescale, we find a reversal 
of this trend, where the observational constraints show a shorter 
$\tau_Q$. This effect could be due to rejuvenation
\citep{Faber:07, Daniel2014}, where a recent episode of star formation
will bias the luminosity-weighted estimates towards the younger
(i.e. lower M/L) component. Since our estimate of $\tau_Q$ is based on
the difference in the stellar age distribution at the 30\% and 70\%,
in luminosity weighting, a recent episode of rejuvenation
can drastically increase
the $\tau_Q$ parameter, however if the episode is of a significant fraction,
it will have an inverse effect, where $\tau_Q$ decreases drastically as
seen in Fig.~\ref{fig:tQ_Age_Params}.

\begin{figure}
    \centering
    \includegraphics[width=85mm]{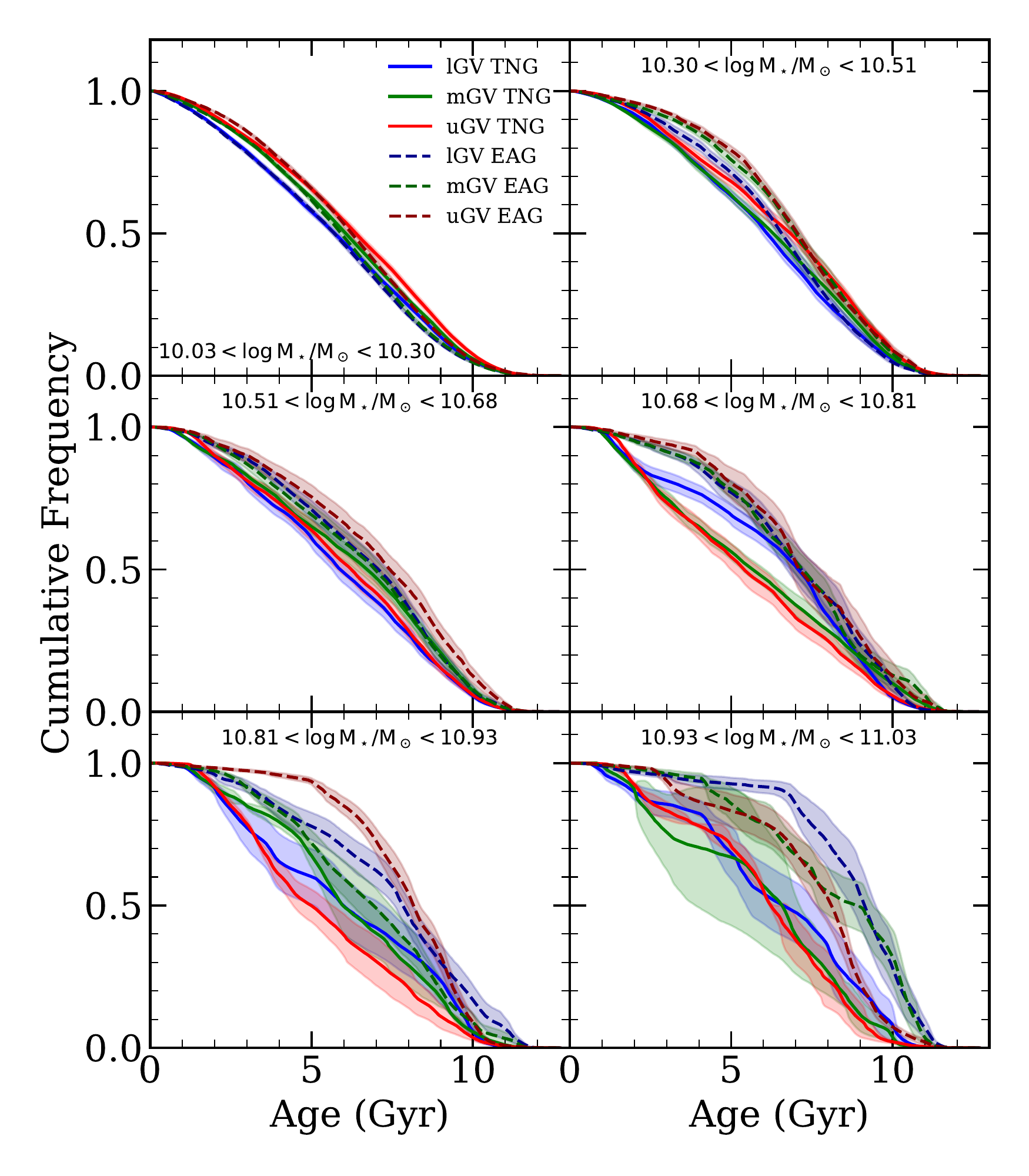}
    \caption{Cumulative frequency of the mass-weighted star formation 
    history of galaxies for different stellar mass bins, as labelled. The blue, 
    green and red data points show results for lGV, mGV and uGV respectively.
    The solid and dashed lines show the mass-weighted star formation
    history in TNG100
    and EAGLE simulations, respectively.}
    \label{fig:Sim_SFH}
\end{figure}

\subsubsection{Simulation Star Formation Histories}
\label{Sec:DnC_SFH}
We have shown that various indicators suggest overquenching or a more
rapid quenching of GV galaxies in
both EAGLE and TNG100 simulations with respect to
the observational evidence provided by SDSS spectra. It is also
interesting to explore the fact that we find a higher fraction of Q
galaxies and lower sSFR in TNG100 with respect to EAGLE, however
EAGLE produces slightly older average ages and shorter quenching
timescales, suggesting earlier and more rapid quenching.
This scenario is illustrated in Fig.~\ref{fig:Sim_SFH}, that shows
the mass-weighted SFHs, as cumulative functions with stellar age, 
where blue, green and red data correspond to the
lGV, mGV and uGV, respectively. The solid and dashed lines are the
results of TNG100 and EAGLE, respectively. At low stellar mass 
($\lesssim 10^{10.68}$M$_\odot$) there is qualitative
agreement between the the two simulations. We note that in these 
stellar mass bins, both simulations produce similar average
ages. Increasing from this stellar mass range, 
EAGLE features a quicker accumulation of stellar
mass at early times compared to TNG100, as shown by the steeper increase
of the cumulative distribution.

The fact that EAGLE galaxies undergo quicker quenching but still
feature higher SFRs shows the nuance and complication we face when
trying to understand galaxy formation and evolution. Normally, a low sSFR,
specially in the GV, is associated to faster quenching. However our results
illustrate that this is not always the case as parameters such as SFR,
are sensitive to young stars. For instance, estimates based on recombination
lines are mostly sensitive to the ionising flux from short-lived O and B type
stars, therefore only encompass the very recent SFH. 

Therefore in this study we see two distinct and different effects 
in EAGLE and TNG100 simulations. EAGLE shows a more rapid quenching
at early times, whereas TNG100 quenches galaxies at later times, both
showing an excess of Q GV
galaxies with respect to the observational constraints,
but with subtle differences. We argue that this difference may be, partly,
caused by the difference in the definition of the black hole seed mass
and halo mass threshold. In EAGLE, the choice is $10^5h^{-1}$M$_\odot$ for the
BH seed and $10^{10}h^{-1}$M$_\odot$ for the halo mass threshold, whereas TNG100
adopts values about 8 times higher in both cases. This increase in the TNG100
simulation is justified to mitigate slow early growth \citep{Pillepich:18},
but it might also delay the onset of AGN feedback into a stage that
removes all chances of a later stage of star formation at the
observed redshift ($z=0.1$), thus overproducing Q galaxies in the GV.

\begin{figure}
    \centering
    \includegraphics[width=80mm, height=90mm]{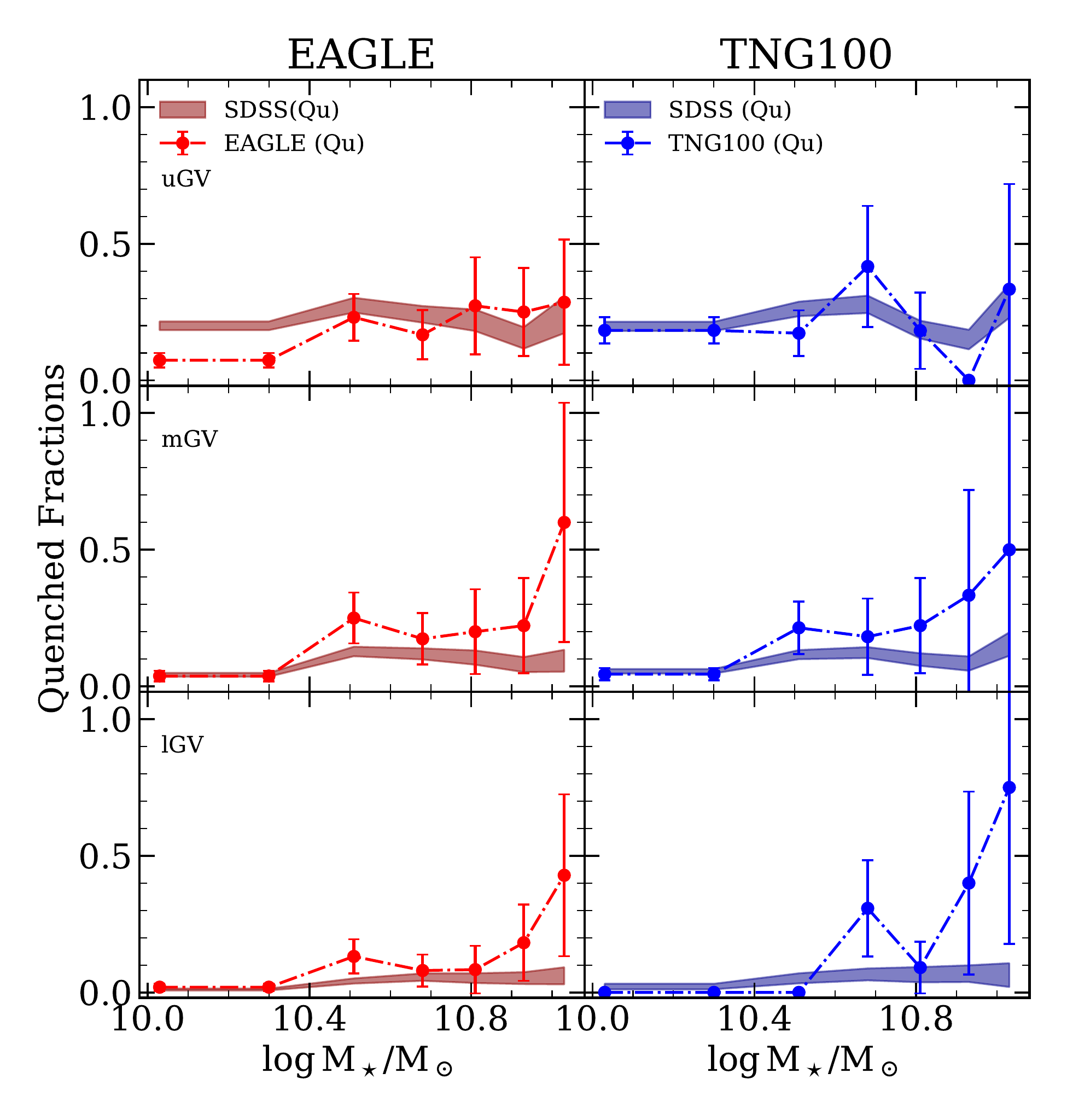}
    \caption{Comparison of the quenched fraction as a 
    function of stellar mass in SDSS (filled) and simulations (dashed).
    The left and right panels show the results for EAGLE and 
    TNG100 data, respectively. 
    Quiescent galaxies in SDSS are selected via the 
    BPT classification ($-1$, corresponding to no emission line detectable). 
    Simulated quiescent galaxies
    are selected using the criteria stated in Tab.~\ref{tab:AGNdef}.
    From bottom to top, each panel
    corresponds to the lGV, mGV and uGV.}
    \label{fig:Frac_zero}
\end{figure}

Fig.~\ref{fig:Frac_zero} shows the quenched fraction in both observation and simulation 
data with respect to all galaxy types (concerning nebular emission).
The SDSS galaxies are flagged as quenched if 
their BPT classification is $-1$ (i.e. no detected emission line).
For the simulations, a quenched galaxy 
follows the criteria shown in Tab.~\ref{tab:AGNdef}, namely: 
$\log\, (\lambda_{\mathrm{Edd}}) < -4.2$ and $\log\, \mathrm{sSFR} \leq -11.0$.
The filled and dashed lines show 
results for SDSS and simulations. Left and right panels
show the results for EAGLE and TNG100, respectively. From bottom to top, 
the panels represent the
lGV, mGV and uGV, evolving from the part of the GV closest to the BC
towards the RS. This figure, along with Fig.~\ref{fig:Sim_SFH} allows
us to provide an explanation for our results.
Both EAGLE and TNG100 show an overall 
agreement with respect to their SDSS counterpart, where
there is an increase  in GV quiescent galaxies
with stellar mass. 
In lGV and mGV, we find a good qualitative
agreement between both simulations and observations. 
The simulations produce a higher quenched fraction with increasing
stellar mass. The exception to this is the uGV, where there is a
constant level of quenched fraction galaxies. Both simulations
are able to reproduce a trend matching the observations.
Both EAGLE and TNG100 simulations also produce 
a similar rate of increase in the quiescent 
population. Note the general low fraction of Q galaxies 
in the GV. This is owing to GV being mostly populated  by 
LINER and Seyfert-type galaxies \citep{Martin:07,Angthopo:19},
further indicating the importance of having the correct subgrid 
physics for AGN feedback in hydrodynamical simulations.

The slower quenching leads to younger stellar components in TNG100
than EAGLE, thus we find younger luminosity-weighted average ages for
TGN100, $t_{\rm fid} = 2.98 \pm 0.08\, $Gyr, than EAGLE,
$t_{\rm fid}= 3.77 \pm 0.33\, $Gyr, (Fig.~\ref{fig:Ave_Age_Params}).
The rapid change in the quenched fraction of GV galaxies and the
higher number of galaxies with zero SFR (Fig.~\ref{fig:sSFR}) provides
an explanation for the observed excess of Q GV galaxies in TNG100
(Fig.~\ref{fig:FracGalSDSS}); with respect to SDSS and EAGLE.
Furthermore, the number of galaxies with zero SFR increases as we go
from lGV to the uGV. This occurs at stellar masses above
$\gtrsim 10^{10.5}$M$_\odot$, where the kinetic feedback is switched on,
thus suggesting
the excess of kinetic feedback at late times might be the reason for
this behaviour.  Note \cite{Nel:18} also explore rejuvenation in
TNG100, however this is prominent at high stellar mass, $\gtrsim
10^{11.0}$M$_\odot$, beyond the stellar mass interval explored in our
study.

\subsection{Subgrid interpretation}
\label{Sec:DnC_improvements}

We explore in this section the details of the subgrid physics
implemented in the simulations that could give rise to the differences
presented above. Given the mass range we are studying, a valid
assumption is to consider AGN feedback as responsible for the
mismatch. Previous work in the literature noted a strong dependence
of quenching within the stellar mass values studied
\citep{KAGN:06,2006LateAGN,2006EarlyAGN}. For galaxies with
stellar mass lower than $10^{9.7}$M$_\odot$, EAGLE finds very low
AGN activity and most quenching is due to stellar feedback or
environment \citep[see, e.g.][]{Crain:15}. Above  $10^{9.7}$M$_\odot$ -- corresponding to the mass
range probed in this study -- EAGLE galaxies quench star formation
via AGN feedback \citep{Bower:17}. Note even within this regime,
EAGLE has two distinct intervals where AGN feedback quenches in different ways.
If the stellar mass is $10^{9.7}<$M$_\star$/M$_\odot<10^{10.3}$, 
EAGLE mimics radio-mode feedback, while more massive galaxies undergo a
rapid increase in the super-massive black hole accretion rate
\citep{Wright:18}. This coincides
with the same mass regime where we see a drastic increase in the
fraction of Q galaxies, with an excess over the observed constraint.
Therefore we can assume the increase in SMBH accretion rate, while
self-regulated, could be too rapid.

In TNG100, the black hole mass and the Eddington ratio has a strong
dependence on how much energy is injected back onto the
environment and in which form. In TNG100, primarily the AGN feedback, 
particularly the kinetic BH-driven winds, have  
been demonstrated to suppress star formation above 
$10^{10}$ M$_\odot$ \citep{Weinberger:17, Nel:18, Terrazas:2020, Dav:2020, Don:20b},
hence the choices in subgrid physics of SMBH seed, growth or
feedback are probably the cause of the discrepancies 
we have seen in this analysis. Specifically at the massive end,
$\gtrsim 10^{10.5}$M$_\odot$, we find a substantial difference between SDSS
and TNG100, where the simulation switches from predominantly 
thermal feedback:
\begin{equation}
        \Delta \dot{E}_{\rm therm} = \epsilon_{\rm f,high}\epsilon_r\dot{M}_{\rm BH}c^2,
\end{equation}
to kinetic feedback:
\begin{equation}
    \Delta \dot{E}_{\rm kin} = \epsilon_{\rm f,kin}\dot{M}_{\rm BH}c^2,
    \label{eq:TNG_Kin_feed}
\end{equation}
thus injecting a higher amount of energy onto the interstellar
medium (ISM) \citep{Weinberger:17}.
Therefore we suspect that the kinetic feedback,
along with the delay in SMBH growth imposed by the higher thresholds
in the choice of seed mass,  may be responsible
for the over-quenching found in GV galaxies 
\citep[see also][who found a similar discrepancy]{Li:2019}.  
At fixed stellar mass, the black hole mass
in TNG100 is greater than the estimates derived from observations \citep{Li:2019},
hence we should expect the energy output, as shown in Eq.~\ref{eq:TNG_Kin_feed},
to be overestimated in this simulation. The overmassive black holes
could also lead to the overquenching in the quasar mode as well, 
as more energy is injected to the surrounding ISM, thus
yielding a higher fraction of Q galaxies in the GV.
Moreover, \citet{Terrazas:2020} found a sharp decline in the SFR
at M$_\star\sim 10^{10.5}$M$_\odot$, further supporting this assumption.
Note this sharp decline in star formation 
could mean TNG100 has a more rapid quenching timescale, 
compared to both EAGLE and SDSS. However the bulk of the quenched
galaxies might reside already in the RS, rather than the GV.
At the same time, the dependence of the quenched fraction
with stellar mass in TNG100 matches the  SDSS data,
when we use a different definition to classify  star-forming and quiescent galaxies
in the general population \citep{Don:20a}.


\section{Conclusions}
\label{Sec:Conc}

We make use of a recent definition of GV galaxies based on 4000\AA\
break strength, which uses SDSS classic spectra, and a probabilistic
approach to separate BC, GV and RS  (see A19, A20 for more details),
to explore two of the latest,
state-of-the-art cosmological hydrodynamical simulations: EAGLE,
(RefL0100N1504), and IllustrisTNG (TNG100-1).
We model the simulated galaxies to obtain
a set of mock spectra without accounting for the effects of dust and by
including the contribution from all the stellar particles in 
a cylinder within 3\,kpc galacto-centric radius to mimic the 3\,arcsec
(diameter) SDSS fibre aperture. We make use of the E-MILES population
synthesis models \citep{eMILES}
to create the spectra. The study projects the
simulated data on the observationally motivated plane, comparing these
simulations with high quality spectroscopic data from the classic SDSS
survey \citep[e.g.][]{DR14}.

The galaxy samples need to be homogenised to avoid selection
biases. Although we homogenise in the stellar mass range
$9\lesssim\log\, \mathrm{M_\star/M_\odot} \lesssim 12$, our analysis
focuses on a narrower region,
$10.03\lesssim\log\, \mathrm{M_\star/M_\odot} \lesssim 11.03$, which
is the stellar mass range obtained by converting velocity dispersion
to stellar mass, see A19 and A20. The velocity dispersion ranges from
$70-250\,$km/s. At higher velocity dispersion the analysis would be
strongly limited by Poisson noise.  Regarding nebular activity, that
allows us to classify the observed spectra, we define a number of
criteria based on the sSFR and the Eddington ratio of SMBH growth to
separate the simulated galaxies into Seyfert AGN, LINER, Q or SF. We
use the global fractions of the homogenised observational sample to
define the constraints, and focus on differences in the trends with
respect to stellar mass. Despite the different subgrid physics
implemented in these two simulations, we find similar constraints in
EAGLE and TNG100 (Tab.~\ref{tab:AGNdef}).

A reasonable agreement is found in general between observations and
simulations, where the simulations are able to produce the bimodal
distribution on the D$_n$(4000) vs stellar mass plane.  Both EAGLE and
TNG100 correctly produce the location of the BC.  However EAGLE
features a RS that appears too high regarding the 4000\AA\ break
strength (by $\Delta$D$_n$(4000)$\sim+$0.1\,dex), while TNG100
produces a RS with an even higher discrepancy
($\Delta$D$_n$(4000)$\sim+$0.2\,dex). Furthermore, 
as previously noted, both simulations produce a RS with a shallower 
gradient compared to SDSS constraints, with TNG100 producing the
flattest RS (Fig.~\ref{fig:GV_D4k_Sim}).
Such a disagreement in the RS is expected to produce a 
similar mismatch in the GV, the main focus of our analysis.
Due to the sparsity of the GV in optical bands,
we only focus on $10.82\%$ EAGLE and $4.91\%$ IllustrisTNG galaxies
of the homogenised samples. Even so, the analysis yields large
constraining capabilities to improve the simulations. We emphasize
that the analysis of GV galaxies provides a fundamental constraint
beyond the {\sl zeroth order} constraints such as the galaxy stellar
mass function, the Tully-Fisher relation, or the mass-metallicity
relation, and focuses narrowly on the subgrid physics that regulates
the SFH of galaxies via feedback. Our analysis, based on the
dust-resilient D$_n$(4000) index provides a robust definition of the
green valley and is complementary to previous comparisons, such as,
e.g., \cite{Tray:15, Tray:17, Nel:18} for the galaxy color,
D$_n$(4000) and SFR vs. galaxy stellar mass planes across the
simulated galaxy populations.

We find both EAGLE and TNG100 overproduce (underproduce) the fraction
of Q (SF) galaxies in the lGV and
mGV\footnote{lGV, mGV and uGV
represent the lower-, mid- and upper- green valley, defined as three
terciles of the 4000\AA\ break strength distribution within bins at
fixed stellar mass} 
in galaxies with stellar mass
$10^{11.0}$M$_\odot$. Moreover, the overproduction of quiescent
GV galaxies is more prevalent in TNG100 than EAGLE
(Fig.~\ref{fig:FracGalSDSS}), supported by the comparison of sSFR.
While both EAGLE and TNG100 produce galaxies with sSFR
marginally greater, by $\sim0.1\,$dex, than the observations, they match 
the overall trend. However, both models  
still show overquenching as there are greater fractions of fully 
quenched galaxies with respect to the observations (Fig.~\ref{fig:sSFR}).
Furthermore, a comparison between simulations shows TNG100 produces more
completely quenched galaxies than EAGLE. Note, by using a 
different definition of the SFR \citet{Don:20b} have found good agreement with
SDSS in the general galaxy population, i.e. not focusing on the GV, to
better than 10 per cent, in the transitional mass scale of
$10^{10-11} \mathrm{M_\odot}$.

Despite EAGLE having a lower fraction of Q GV galaxies, we find they
produce overall older GV galaxies, both in luminosity and mass
weighting, and undergo more rapid quenching at high stellar mass than
TNG100.  In comparison to the observations, both simulations produce
older luminosity-weighted stellar ages. 
EAGLE and TNG100 produce up to $3.5\,$Gyr and up
to $2.8\,$Gyr older galaxies, respectively, with a dependence on
stellar mass.  EAGLE yields a steeper gradient of the correlation
between luminosity-weighted average age and stellar mass, whereas
TNG100 is able to produce a slope that roughly matches the SDSS-based
constraints. The luminosity-weighted quenching timescale ($\tau_Q$)
shows a more extended transition time in simulations with respect to
the SDSS data, a result that is reversed in mass-weighted $\tau_Q$
(Fig~\ref{fig:tQ_Age_Params}). However both simulations are able to
reproduce the observational trend, more specifically a decrease in the
mass-weighted quenching timescale with respect to stellar mass.

Finally, at the massive end, $10^{10.5-11.0}$M$_\odot$, EAGLE GV
galaxies undergo more rapid quenching compared to TNG100
(Fig.~\ref{fig:Sim_SFH}).  Both simulations show signs of
overquenching, where, at higher stellar mass
$10^{10.5-11.0}$M$_\odot$, there is a higher quenched fraction with
respect to SDSS.  While both over-produce quenched galaxies, we find a
larger discrepancy between TNG100 and SDSS, up to $0.62$ (fractional excess), 
compared to EAGLE and SDSS, with a difference up to 
$0.50$ (Fig.~\ref{fig:Frac_zero}). This shows that while TNG100 
galaxies tend to  quench at later times than EAGLE, 
they also quench more efficiently on
the GV. This suggests that EAGLE allows for later episodes of star formation, 
when measured using the instantaneous SFR, which gives results that appear
to be more in agreement with constraints from the SDSS sample
(Fig.~\ref{fig:sSFR}). Multiple studies have noted the strong quenching 
nature of the kinetic black hole-driven winds \citep{Terrazas:2020, Dav:2020}, 
therefore we ascribe this difference to AGN feedback, which sterilise the ISM of
TNG100 galaxies, resulting in an excess of quiescent galaxies on the
GV. This paper illustrates the power of the green valley -- defined by use of the 4000\AA\ break strength -- as a key
laboratory where feedback prescriptions can be put to the test on
state-of-the-art simulations of galaxy formation.

\section*{Acknowledgements}
JA is supported by the UK Science and Technology 
Facilities Council (STFC).
We acknowledge support from the Spanish Ministry of Science, Innovation
and Universities (MCIU), through grants PID2019-104788GB-I00 (IF), 
and PID2019-107427GB-C32 (IGdlR).
This work has also been supported by the IAC project TRACES, 
partially funded through the state budget and the regional budget
of the Consejer\'\i a de Econom\'\i a, Industria, Comercio y
Conocimiento of the Canary Islands Autonomous Community.

\section*{Data Availability}
This project is fully based on publicly available data from SDSS, EAGLE,
and IllustrisTNG project. The data used for this project are available
on request.


\bibliographystyle{mnras}
\bibliography{SimGV} 

\appendix

\section{Comparing homogenised samples}
\label{sec:Frac_Match}

\begin{figure}
    \centering
    \includegraphics[width=70mm,height=65mm]{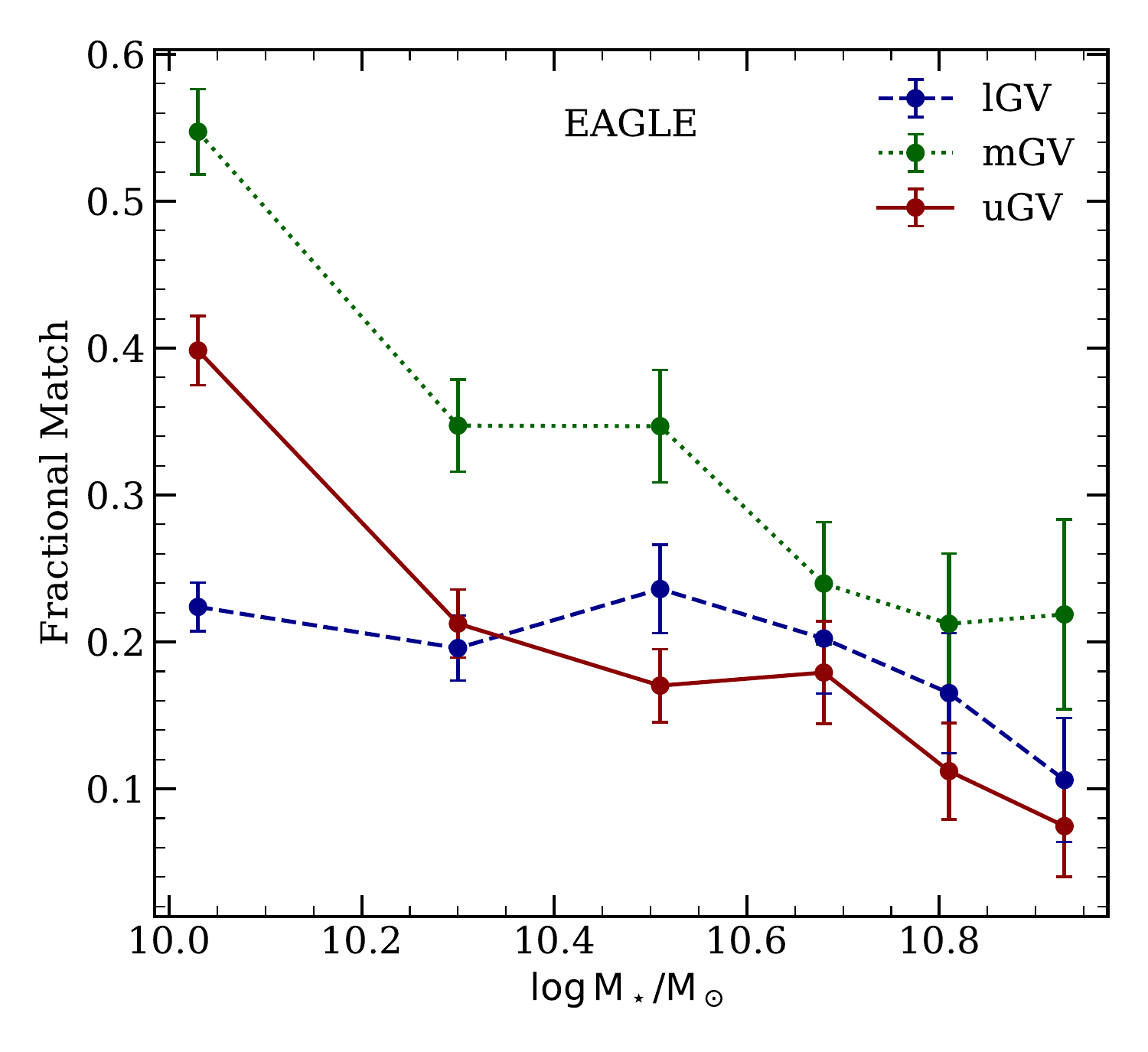}
    \caption{Fraction of matching SDSS galaxies in the EAGLE homogenised 
    simulation for lGV (blue), mGV (green) and uGV (red) galaxies. The
    error bars assume Poisson noise. The fractions are similar
    when selecting GV galaxies from the homogenised TNG100 sample.}
    \label{fig:frac_match}
\end{figure}

Since the homogenisation procedure results in different SDSS
subsamples when comparing EAGLE and TNG100 simulations, we want to assess
the level of overlap between these two pairs of samples. In each pair
(i.e. either SDSS-EAGLE or SDSS-TNG100) we follow the methodology laid
out in A19 and A20 to define the uGV, mGV and lGV subsamples within
each stellar mass bin -- defined as the lower-, mid-, and upper-GV,
respectively, which are meant to map three regions, defined
by splitting GV into three terciles of the 4000 \AA\ break strength for
individual stellar mass bins, that follow the
transition from BC into RS.
In each subset, for instance the mid green valley, mGV within 
the 10.30--10.51 (log) stellar mass bin, we cross-correlate the
SDSS-EAGLE and SDSS-TNG100, and identify the number of galaxies in both
sets, expressing this number as a fraction with respect to the total
in each bin. Fig.~\ref{fig:frac_match} shows a comparison of these
fractions as a function of stellar mass, where the blue, green and red
data points refer to lGV, mGV and uGV, respectively. The top and
bottom panels show the fractional match for EAGLE and TNG100,
respectively.

The error bars have been obtained assuming Poisson noise in the
count of overlapping galaxies. The fractional 
match decreases with increasing stellar mass in all subsets.
This is due to  (i) the independent homogenisation of EAGLE and TNG100, and
(ii) the selection of  GV galaxies is probability-based, therefore we do not find 
a unique solution, and is subject to Poisson noise.
The difference in fractional match shown in Fig.~\ref{fig:frac_match}
will lead to differences in the retrieved SDSS parameters for
EAGLE and TNG100 sample, as seen below,
specially in Section \ref{sec:Ave_time} and
\ref{sec:Quench_time}.

\label{lastpage}

\end{document}